\documentclass{aa}  

\usepackage{graphicx}
\usepackage{txfonts}
\usepackage{natbib}
\usepackage{ulem}

\begin{document}
  
  \title{Protostellar collapse: A comparison between SPH and AMR calculations.}
  
   \author{B. Commer\c con
           \inst{1,2,3}
           ,
           P. Hennebelle\inst{3}
 	  ,
	  E. Audit\inst{2}
	  ,
 	  G. Chabrier\inst{1}
	  \and
	  R. Teyssier\inst{2}
          }
   \offprints{B. Commer\c con}

   \institute{\'Ecole Normale Sup\'erieure de Lyon, Centre de recherche Astrophysique de Lyon (UMR 5574 CNRS), 
46 all\'ee d'Italie, 69364 Lyon Cedex 07, France\\
              \email{benoit.commercon@cea.fr}
         \and 
	 Laboratoire AIM, CEA/DSM - CNRS - Universit\'e Paris Diderot,
DAPNIA/SAp, 91191 Gif sur Yvette, France 
         \and
             Laboratoire de radioastronomie millim\'etrique (UMR 8112 CNRS), \'Ecole Normale Sup\'erieure et Observatoire 
de Paris, 24 rue Lhomond, 75231 Paris Cedex 05, France\\
             }

   \date{Received September 3, 2007; accepted January 16, 2008}

  \abstract {The  development of parallel  supercomputers allows today
      the  detailed study  of the  collapse and  the  fragmentation of
      prestellar   cores    with   increasingly   accurate   numerical
      simulations.    Thanks  to   the   advances  in   sub-millimeter
      observations, a wide range of observed initial conditions enable
      us to study the different  modes of low-mass star formation. The
      challenge for the simulations  is to reproduce the observational
      results.}  {Two main numerical  methods, namely AMR and SPH, are
      widely used  to simulate the  collapse and the  fragmentation of
      prestellar cores. We compare thoroughly these two methods within
      their standard framework.}  {We use  the AMR code RAMSES and the
      SPH code DRAGON.   Our physical model is as  simple as possible,
      and  consists  of  an  isothermal  sphere  rotating  around  the
      $z$-axis.  We  first study the conservation  of angular momentum
      as a function of the  resolution.  Then, we explore a wide range
      of   simulation  parameters  to   study  the   fragmentation  of
      prestellar cores.}  {There seems to be a convergence between the
      two  methods, provided  resolution in  each case  is sufficient.
      Resolution criteria  adapted to our physical cases,  in terms of
      resolution per  Jeans mass, for  an accurate description  of the
      formation  of protostellar  cores are  deduced from  the present
      study.   This  convergence is  encouraging  for  future work  in
      simulations   of   low-mass   star  formation,   providing   the
      aforementioned  criteria  are  fulfilled.}  {}  \keywords{Stars:
      formation - Methods : numerical - hydrodynamics}

\titlerunning{Protostellar collapse: A comparison between SPH and AMR calculations.}
\authorrunning{B. Commer\c con et al.}
   \maketitle

\section{Introduction}
Star formation is  known for being the place  of extreme variations in
length and density scales. Although  it is established that stars form
in dense cores, the non-linear evolution makes it difficult to perform
accurate  calculations of  the  collapse and  the  fragmentation of  a
prestellar core. The star formation  process is the outcome of complex
gas dynamics involving non-linear interactions of gravity, turbulence,
magnetic  field and  radiation.   Early theoretical  pioneer works  by
\cite{Larson_1969},  \cite{Penston_1969} or \cite{Shu_1977}  are among
the  many illustrations of  the high  complexity of  the gravitational
collapse.  Recently,  \cite{Klein_2007} pointed out  that developing a
theory for low-mass star formation remains one of the most elusive and
important  goals  of   theoretical  astrophysics.   The  computational
challenge stems  from the  fact that star  formation occurs  in clouds
over  many  orders  of   magnitude  in  spatial  and  density  scales.
Following  the gravitational  collapse while  resolving  precisely the
Jeans length, which scales as $\lambda_\mathrm{J} \propto \rho^{-1/2}$
for  an   isothermal  gas,  is   a  major  difficulty   for  numerical
simulations.

Different  approaches  are  used   to  study  star  formation  through
numerical simulations and include  more and more detailed physics. One
key question resides  in the validation of the  numerical methods used
to study low-mass star  formation.  Nowadays, two completely different
numerical method are used with sufficient accuracy:

\begin{enumerate}
  \item AMR: Adaptive Mesh Refinement method for Eulerian grids
  \item SPH:  Smoothed Particle Hydrodynamics method  for a Lagrangian
  approach.
\end{enumerate}

No systematic  comparison between the  two methods has been  done with
low-mass  star formation  calculations.  However,  a lot  of numerical
works  have  been  carried  out  and  some of  them  are  common  test
calculations for  convergence testing and  intercode comparisons.  The
most famous  model was first calculated by  \cite{Boss_1979} and since
then, it  has been  recalculated by several  authors with  even higher
spatial           resolution           \cite[e.g][]{bate_burkert-1997,
Truelove_1998,Kitsionas_2002,Arreaga_2007}.   The   SPH  approach  has
generated a  lot of  detailed investigations on  the influence  of the
number          of           particles          and          neighbors
\citep{Lombardi_1999,rasio-1999,Attwood_2007},    and   criteria   for
numerical  convergence   have  been  extracted   from  these  studies.
\cite{Nelson_2006} performed a large investigation of the influence of
these parameters on disk  fragmentation, and concluded that the better
the  resolution  the   later  the  fragmentation.   \cite{Dehnen_2001}
investigated  the optimal gravitational  force softening  necessary in
three-dimensional $N$-body  codes.  \cite{bate_burkert-1997} provide a
minimum resolution criterion for SPH calculations with self-gravity to
accurately model fragmentation.  Less studies have been performed with
AMR    since    AMR   codes    have    become   available    recently.
\cite{Truelove_1997} give an empirical  criterion for the Jeans length
resolution   in   AMR  calculations   to   avoid  spurious   numerical
fragmentation.

There   are  not  much of    direct  comparison  between SPH  and  AMR
calculations. Comparison  in  the context of  cosmological simulations
has   been done through the Santa   Barbara Cluster Comparison Project
\citep{Santa_Barbara_1999}.        \cite{Fromang_2006}  compares quite
successfully AMR hydrodynamical collapse calculations with the ones of
\cite{Hosking_2004}, using the SPH method.\\

In the present paper, we  compare thoroughly the two approaches in the
context  of low-mass prestellar  core formation.   We stress  that the
main goal of  this paper is to investigate  whether convergence can be
achieved between the two methods.  We have conducted calculations over
a wide range of numerical resolution parameters, in order to study the
dependency  of  angular  momentum  conservation and  fragmentation  on
physical and numerical initial  conditions.  We then derive resolution
criteria necessary to describe accurately prestellar core formation.


The paper  is organized  as follows: in  \S2 we briefly  introduce our
collapse  model.   In \S3,  we  present the  two  codes  used for  our
comparative study as well as  our initial numerical conditions and the
criteria we fulfill to resolve gravitational collapse.  The problem of
angular momentum conservation  is examined in detail \S4.   In \S5, we
tackle  the fragmentation  issue  and explore  the  dependency of  the
results  on the numerical  parameters. First,  we study  the numerical
convergence of AMR and  SPH calculations separately.  Then, we compare
the respective converged calculations.  This convergence study is done
for  different test  cases. In  section 6  we conclude  this  paper by
deriving for each method  empirical required numerical criteria for an
accurate  description  of  gravitational  collapse  and  fragmentation
adapted to our test cases.\\

The convention in this paper  is to call "particles" the SPH particles
and "cells" the AMR cells in order to avoid confusion.

\section{Definitions of the test cases}

\subsection{Model}
To  make comparison  between  codes easier,  we  adopt simple  initial
conditions,   similar   to    those   chosen   in   previous   studies
\citep[e.g.][]{Boss_1979,bate_burkert-1997}.      We    consider    an
uniform-density  sphere  of molecular  gas  of  initial radius  $R_0$,
rotating  around  the  $z$-axis   with  an  uniform  angular  velocity
$\Omega_0$, in order  to minimize the loss of  angular momentum due to
friction.   We fix  the cloud  mass at  $M_0 =  1$ M$_{\sun}$  and the
temperature at 10 K.  For  a mixture of molecular hydrogen, helium and
heavy  elements, this  corresponds  to an  isothermal  sound speed  of
$C_\mathrm{0}   \sim  0.19   $  km.s$^{-1}$.    For  the   case  where
fragmentation  occurs,  we  use  a  $\mathrm{m}=2$  azimuthal  density
perturbation.

The  initial  energy  balance  of  our  model  is  determined  by  two
dimensionless  parameters  corresponding  to  the  ratio  between  the
thermal energy and the gravitational energy
\begin{equation}
\alpha = \frac{5}{2}\frac{R_0kT}{GM_0\mu m_\mathrm{H}},
\end{equation}

 and to the ratio of the rotational and the 
gravitational energy
\begin{equation}
\beta = \frac{1}{3}\frac{R_0^3\Omega_0^2}{GM_0}.
\end{equation}

Since we  use a constant initial  mass of 1 M$_{\sun}$  and a constant
temperature,  changing one  of  the two  parameters, namely  $\alpha$,
gives  the  sphere radius  $R_0$.   The  higher  $\alpha$, the  larger
$R_0$. The angular velocity is given by the parameter $\beta$.

\subsection{The barotropic equation of state}

In order to mimic the thermal behaviour of a star-forming gas, we use
a barotropic equation of state
\citep[cf.][]{Bonnell_1994}. \citet{Tohline_1982} and
\citet{Masunaga_Inutsuka_2000} showed that the core follows closely a
barotropic equation of state, providing a good approximation without
resolving radiative transfer. We use
\begin{equation}
\frac{P}{\rho} = C_\mathrm{s}^2 = C_0^2\left[1+\left(\frac{\rho}{\rho_c}\right)^{2/3}\right],
\label{baro}
\end{equation}
where  $C_\mathrm{s}$  is the  sound  speed  and  $\rho_c =  10^{-13}$
g.cm$^{-3}$  is   the  critical  density  which   corresponds  to  the
transition    from    an   isothermal    to    an   adiabatic    state
\citep{Larson_1969}.

At  low  densities,  $\rho  \ll \rho_\mathrm{c}$,  $C_\mathrm{s}  \sim
C_0=0.19 $ km.s$^{-1}$. The molecular gas is able to radiate freely by
coupling thermally to the dust  and therefore remains isothermal at 10
K.  At  high densities  $\rho > \rho_\mathrm{c}$,  we assume  that the
cooling  due to  radiative transfer  is trapped  by the  dust opacity.
Therefore,  $P\propto \rho^{5/3}$  which corresponds  to  an adiabatic
monoatomic  gas with  adiabatic exponent  $\gamma =  5/3$.   Note that
molecular hydrogen behaves like a monoatomic gas until the temperature
reaches  several  hundred  Kelvin,  since the  rotational  degrees  of
freedom are  not excited  at lower temperatures,  and hence  $\gamma =
5/3$       is      the       appropriate       adiabatic      exponent
\citep{Whitworth_Clarke_1997,Masunaga_Inutsuka_2000}.

\section{Numerical methods and initial conditions}

\subsection{AMR: Adaptive Mesh Refinement}

\subsubsection{A brief history}

The  Adaptive Mesh  Refinement method  is  one of  the most  promising
numerical  methods to solve  the fluid  equations.  The  technique was
first  introduced in  \citet{berger_oliger_1984}. Originally,  the AMR
method  was an  Eulerian hydrodynamical  scheme, with  a  hierarchy of
nested grids covering high resolution  regions of the flow. This first
AMR structure, called ``patch-based AMR'', consists of building blocks
of the computational grid as  rectangular patches of various sizes. An
alternative  method  was  proposed \citep[i.e.][]{khokhlov_1998},  the
``tree-based'' AMR,  where the parent cells are  refined into children
cells  on  a cell-by-cell basis.   These adaptive mesh  structures are
coupled    with   grid-based    fluid   dynamics    schemes   handling
high-resolution shock capturing.  Nowadays, high order Godunov methods
appear  to be  amongst  the best  schemes  to capture  discontinuities
within             only             a            few             cells
\citep[e.g.][]{teyssier-2002,Matsumoto_2003,Ziegler_2005,Fromang_2006}.

\subsubsection{The RAMSES code}

In this paper, we use the AMR code RAMSES \citep{teyssier-2002}, which
integrates the  ``tree-based'' data structure  allowing recursive grid
refinements. RAMSES uses a  second order Godunov hydrodynamical scheme
coupled with a gravity solver.  Furthermore, it has the possibility to
use  variable  timesteps at  each  refinement  level. Concerning  time
integration,  RAMSES  uses   a  second-order  midpoint  scheme,  where
positions and  velocities are  updated by a  predictor-corrector step.
Recently,  an  ideal MHD  version  of  RAMSES  has been  developed  by
\cite{Fromang_2006}.\\

The Godunov  hydrodynamical solver is able  to capture discontinuities
with a  high precision level. The  equations solved in  RAMSES are the
Euler equations in their conservative form
\begin{equation}
\label{masse} \frac{\partial \rho}{\partial t} + \nabla\cdot \left[\rho\textbf{u} \right] = 0 ,
\end{equation}

\begin{equation}
\label{mvt} \frac{\partial \rho \textbf{u}}{\partial t} + \nabla \left[\rho \textbf{u} 
\otimes \textbf{u} + P \mathbb{I} \right] = -\rho\nabla\Phi,
\end{equation}

\begin{equation}
\label{nrj} \frac{\partial  \mathrm{E}}{\partial t} + \nabla \left[\textbf{u}\left(\mathrm{E} + P   \right)\right] = -\rho
 \textbf{u}\cdot \nabla \Phi.
\end{equation}
where $P$ is the gas pressure ($\mathbb{I}$ is the identity matrix), $\rho$ the
density,  $\textbf{u}$  the velocity,  $\mathrm{E}$  the total  energy
density  and  $\Phi$  the  gravitational  potential.   The  system  of
equations   is  closed   with   the  barotropic   equation  of   state
(\ref{baro}).

One of  the main  advantages of solving  the Euler equations  in their
conservative form is  that no energy sink due  to numerical errors can
alter the  flow dynamics (ignoring  the source terms due  to gravity).
Equations  \ref{masse},  \ref{mvt} and  \ref{nrj}  are  solved with  a
Lax-Friedrich Riemann solver,  known to be one of  the most simple and
robust  scheme.   In  Appendix   \ref{note_diff},  we  report  on  the
influence  of  using  the  Roe  solver which  is  less  diffusive  but
considerably more complex and numerically expensive.

The  timestep is  determined independently  for each  refinement level
$\ell_i$, using standard  stability constraints for the hydrodynamical
solver.  Each level $\ell_i$ evolves according to its own timestep.

\subsubsection{Initial conditions for RAMSES}
One practical limitation  of AMR codes is the  use of Cartesian grids.
RAMSES works  with a cubic volume,  so that a part  of the calculation
box is lost when we describe  a sphere. The outer region of the sphere
is also at a 10 K  temperature but is 100 times less dense. Therefore,
the outer  gas has no effect on  the dynamics of the  sphere since the
two parts are well separated.  The sphere radius is equal to a quarter
of the box length in order to minimize border effects.

\subsection{SPH: Smoothed Particles Hydrodynamics}

\subsubsection{A brief overview}

SPH  is the  most popular  fully  Lagrangian method  used to  describe
gravitational collapse because of its  simplicity for 3D codes and its
versatility  to incorporate  self-gravity. SPH  was first  designed to
simulate    nonaxisymmetric   phenomena   for    astrophysical   gases
\citep{Lucy_1977,Gingold_Monaghan_1977}.  This method  is easy to work
with and can give rapidly reasonably accurate results. SPH is economic
in  handling hydrodynamical flows  that have  near empty  regions.  It
does not need a grid to calculate spatial derivatives, but consists of
a set  of discrete  particles describing the  state of the  fluid. The
spatial  derivatives  are   found  by  analytical  differentiation  of
interpolation formulae. SPH particle $i$  should not be perceived as a
real  fluid element,  but as  a mathematical  entity  with coordinates
$\textbf{r}_i$, velocity  $\textbf{v}_i$, mass $m_i$  (i.e.  $m$ since
all  particles have  the same  mass in  the present  calculations) and
thermal energy  $e_i$.  The  evolution of the  fluid is  determined by
following  the  motion  of  the  particles,  under  the  influence  of
interparticle  forces   which  represent  the   effects  of  pressure,
viscosity (see below) and self-gravity.

The main  advantage of SPH  is its strict  Galilean-invariant property
 and  its simplicity.   Resolution elements  are then  concentrated in
 high density  regions in  SPH methods .   The standard  SPH formalism
 uses  artificial viscosity for  the hydrodynamics.   Some alternative
 formalism    such    as     Godunov    SPH    has    been    proposed
 \citep[e.g.][]{Inutsuka_1994} in order to avoid the use of artificial
 viscosity, but these  methods are not yet mature.   SPH has been used
 by       several      authors       to       study      fragmentation
 \citep{Bonnell_1994,bate_burkert-1997,Goodwin_2004,Hennebelle_2004}.

\subsubsection{Hydrodynamical method for DRAGON}\label{sph}

We       use       the       standard      SPH       code       DRAGON
\citep{Turner_1995,Goodwin_2004}, i.e in its most simple version.
In   standard  SPH,   the  integral   interpolant  for   the  variable
$A(\textbf{r}_i)$  is approximated by  a summation  interpolation over
the particle's nearest neighbors:
\begin{equation}
  \label{eq:sum}
  A_s(\textbf{r}_i) = \sum_j m_j\frac{A(\textbf{r}_j)}{\rho
  (\textbf{r}_j)}W(|\textbf{r}_i-\textbf{r}_j|,h_{ij}),
\end{equation}
where  $A(\textbf{r}_j)$ is  the value  associated with  particle $j$,
$h_{ij}=(h_i+h_j)/2$  and $h_i$  is the  adaptive smoothing  length of
particle $i$, defined such that  the particle kernel volume contains a
constant    mass,    i.e.    a    constant    number   of    neighbors
$N_\mathrm{N}$.   The   interpolation    mass   is   then   given   by
$mN_\mathrm{N}$.

We use a standard artificial viscosity scheme \citep{gingold_monaghan_1983}:
\begin{equation}
\Pi_{ij} = \left\{
\begin{array}{ccc}
\left(-a_1 C_{0,ij}\mu_{ij} + a_2 \mu^2_{ij}\right)/ \rho_{ij} & & \mbox{if } \textbf{u}_{ij}\cdot\textbf{r}_{ij}\leq 0,\\
0 & & \mbox{otherwise},
\end{array}\right.
\label{visc_sph}
\end{equation}
with
\begin{equation}
 \mu_{ij} = \frac{h_{ij}\textbf{u}_{ij}\cdot\textbf{r}_{ij}}{\textbf{r}_{ij}^2+ a_3^2},
\end{equation}
where    $\textbf{u}_{ij}=     \textbf{u}_{i}    -    \textbf{u}_{j}$,
$\textbf{r}_{ij}= \textbf{r}_{i} -\textbf{r}_{j}$ and ${C}_{0,ij}$ and
$\rho_{ij}$ denote arithmetic means  of the isothermal sound speed and
density of the particles $i$ and $j$. The free parameters $a_1$, $a_2$
and $a_3$ regulate the strength of the viscosity. In our case, we have
the  combination $a_1  = 1$,  $a_2 =  2$ and  $a_3 =  0.1h_{ij}$. This
artificial  viscosity  is the  subject  of  several discussions  where
authors   suggest  alternative   artificial  viscosity.    One  common
improvement     is    to     use     a    time-dependent     viscosity
\citep{Morris_Monaghan_1997}.  We   look  at  the   influence  of  the
viscosity scheme in Appendix \ref{app_visc}.

The gravitational  force between  a pair of  particles obeys  a simple
 inverse-square law,  unless the particles are very  close. Under this
 circumstance, the gravity  force has to be softened  to avoid violent
 two-body  interaction.   According  to \cite{bate_burkert-1997},  the
 gravitational softening length should be the same as the hydrodynamic
 smoothing length. Calculation of the gravitational acceleration of an
 SPH particle  is speeded up  using an octal Spatial  Tesselation Tree
 (STT) \citep{Hernquist_1987} and  accounts for the quadrupole moments
 of the  mass distributions.  DRAGON benefits  from the implementation
 of  ``sink particles'' creation  \citep{Bate_1995}, used  to continue
 the  calculations  without  resolving  processes on  extremely  short
 time-scales.   Last  but  not  least, DRAGON  uses  multiple-particle
 timesteps.

Finally,  we allow  a variation  of  the number  of neighbors  $\Delta
  N_\mathrm{N}$  less   than  10\%   of  $N_\mathrm{N}$  in   our  SPH
  calculations,  i.e $\Delta  N_\mathrm{N}=5$  when $N_\mathrm{N}=50$.
  \cite{Attwood_2007}  shows that  the smaller  $\Delta N_\mathrm{N}$,
  the less diffusive is  SPH. Ideally, $\Delta N_\mathrm{N}$ should be
  set  to  0.    We  report  on  the  influence   of  setting  $\Delta
  N_\mathrm{N}=0$ for local  angular momentum conservation in Appendix
  \ref{app_visc}.

\subsubsection{Initial conditions for DRAGON}

We  use the  method originally  presented by  \cite{Whitworth_1995} to
obtain an initial particle  distribution which consist in settling the
positions   of  the   particles   randomly  settled   by  using   only
hydrodynamical  forces   over  a  few   timesteps.   Another  approach
sometimes used is to take initial hexagonal-close-paced lattice of SPH
particles  to  generate  initial  conditions.  However,  standard  SPH
calculations  start  from  noisy  initial  particle  distributions  at
present time  \citep[e.g.][]{Arreaga_2007}. Note  that we do  not need
intercloud  and external  particles  to confine  the  ones within  the
sphere since our model is initially far from equilibrium.

\subsection{The Jeans criterion in numerical codes}

\subsubsection{Refinement criterion for the AMR method}

Our refinement criterion is based on the Jeans length resolution which
is necessary to treat  accurately gravitational collapse.  We impose a
minimum   number   of   points   $N_\mathrm{J}$   per   Jeans   length
$\lambda_\mathrm{J}$.  The  cells' dimensions  must be smaller  than a
constant fraction of  the local Jeans length.  The  dimension of cells
belonging     to     the      $\ell_i$     refinement     level     is
$L_\mathrm{box}/2^{\ell_i}$,  where $L_\mathrm{box}$  is  the physical
length of the simulation box.  The mesh is locally refined in order to
satisfy the local Jeans criterion:
\begin{equation}
\frac{L_\mathrm{box}}{2^{\ell_i}}< \frac{\lambda_\mathrm{J}}{N_\mathrm{J}}.
\end{equation}

\cite{Truelove_1997}  defined a minimum  resolution condition  for the
validity of grid-based simulations aimed at modeling the collapse of a
molecular  cloud  core,  namely  $N_\mathrm{J} >  4$.  This  condition
ensures  that the  collapse is  of physical  rather than  of numerical
origin.

\subsubsection{Jeans length description with a SPH code}

In standard  SPH, the resolution in  mass is fixed and  thus the Jeans
length  resolution   deteriorates  with  increasing   density  for  an
isothermal gas.  The minimum resolvable  mass must then be larger than
the  interpolation  mass.   \cite{bate_burkert-1997} showed  that  the
behaviour  of  a Jeans-mass  clump  of gas  with  radius  $\sim h$  is
dominated  by the numerical  implementation.  \cite{bate_burkert-1997}
take the smallest mass that can  be resolved in SPH calculations to be
equal to the mass of $\sim 2N_\mathrm{N}$ particles. According to this
criterion,  we  can  determine  an  initial number  of  SPH  particles
necessary to solve the Jeans length in the simulations.

The          Jeans          mass         is          $M_\mathrm{J}\sim
6\,G^{-3/2}\rho^{-1/2}C_\mathrm{s}^3$ and  the minimum resolvable mass
is $M_{res}=m  N_\mathrm{N}$. Hence, we  can define a  Jeans condition
corresponding  to the  minimum  value of  $C_\mathrm{s}^3\rho^{-1/2}$,
given   by   the    barotropic   equation   of   state   (\ref{baro}),
i.e. $2^{3/2}C_0^{3}\rho_c^{-1/2}$:
\begin{equation}
m < m_{max} \sim \frac{2^{3/2} 6\, C_0^{3}}{2N_\mathrm{N}
G^{3/2}\rho^{1/2}_c} \sim \frac{5.35 \times 10^{-3}}{N_\mathrm{N}} \mathrm{ M}_{\sun} .
\label{jeans_sph}
\end{equation}
Considering an initial spherical  mass $M_0=1$ M$_{\sun}$, the initial
number of particles $N_\mathrm{p}$  has to satisfy $N_\mathrm{p} > M_0
/m_\mathrm{max} \sim 9300 $ if $N_\mathrm{N}=50$. This is the critical
number of particles used in  SPH calculations to study the collapse of
a dense core.  We have  in that case exactly $2N_\mathrm{N}$ (i.e. two
resolution    elements)   particles    per   critical    Jeans   mass.
\cite{Hubber_2006} shows that with this numerical resolution, standard
SPH will capture fragmentation which is genuine and resolved.

As mentioned before, the mass  resolution is fixed in standard SPH. In
term of Jeans mass, the  resolution is therefore high at the beginning
of the simulation  and decreases when the density  increases up to the
critical density.  It  is nevertheless instructive to have  a means of
comparing the SPH and AMR  resolution. We therefore define for the SPH
the    parameter    $N_\mathrm{J}$    such    that    $N_\mathrm{J}^3=
M_\mathrm{J}/M_{res}$ is  the number  of resolution element  per Jeans
mass.    This   number   is   computed   at   the   critical   density
$\rho_\mathrm{c}$, i.e.   the most unfavorable  case for the  SPH.  In
the  dense   core  where  $\rho  >   \rho_\mathrm{c}$,  the  parameter
$N_\mathrm{J}$ enables  us to compare  the resolution achieved  by SPH
and  AMR, respectively.   Using particle  splitting refinement  in SPH
\citep[e.g.][]{Kitsionas_2002} would  improve its resolution.  Indeed,
particle splitting is an economic way to increase the local resolution
and  thus   to  avoid  violating  the  Jeans   condition  in  collapse
simulations.  However,  as mentioned in  the introduction, the  aim of
the  present paper  is to  compare the  AMR and  the SPH  within their
standard implementation.

\section{Free-fall time and angular momentum conservation}\label{sec_momang}

We start by comparing the global properties of the collapse in the two
codes in the simple case of an uniform-density sphere collapse with no
perturbation. We  look at the  collapse time, the accretion  shock and
finally the angular momentum conservation.
 
We  carried out  a first  set of  simulations within  a wide  range of
resolution  parameters. The  initial sphere  is set  up  by parameters
$\alpha =  0.65$ corresponding to an  initial radius $R_\mathrm{0}=9.2
\times  10^{16}$ cm and  a density  $\rho_0\sim 6.02  \times 10^{-19}$
g.cm$^{-3}$.   The corresponding  free-fall time  is  t$_\mathrm{ff} =
(3\pi / 32 G \rho_0)^{1/2} \sim 86 $ kyr.

\begin{table}[htb]
\caption{Summary  of the different  simulations for  the case  with no
rotation (left  table: SPH; right table: AMR).  $N_\mathrm{i}$ for the
AMR calculations gives  us the number of cells  describing the initial
sphere.  } \centering
\begin{tabular}{cc}
\begin{tabular}{cccc}
\hline
\hline
$N_\mathrm{p}$    & $N_\mathrm{N}$ & $N_\mathrm{J} $ & t$_0$ (kyr) \\
\hline
$5 \times 10^3$  & 50    & 1.86 & 108.6 \\
$1\times 10^4$   & 50    & 2.33 & 102   \\
$5\times 10^4$   & 50    & 4.   & 94.1  \\
$2\times 10^5$   & 50    & 6.35 & 91.5  \\
$5\times 10^5$   & 50    & 8.61 & 90.6  \\
\hline
\end{tabular} 
\begin{tabular}{cccc}
\hline
\hline
 $\ell_\mathrm{min}$   & $N_\mathrm{i}$ & $N_\mathrm{J}$  & t$_0$ (kyr)\\
\hline 
5   &   2 145    & 10 & 109 \\
6   &  17 160    & 10 & 98  \\
7   & 137 260    & 10 & 95  \\
\hline
\end{tabular}
\end{tabular}
\label{sum_ff}
\end{table}

\begin{figure}[h]
  \centering
  \includegraphics[width=8cm,height=6cm]{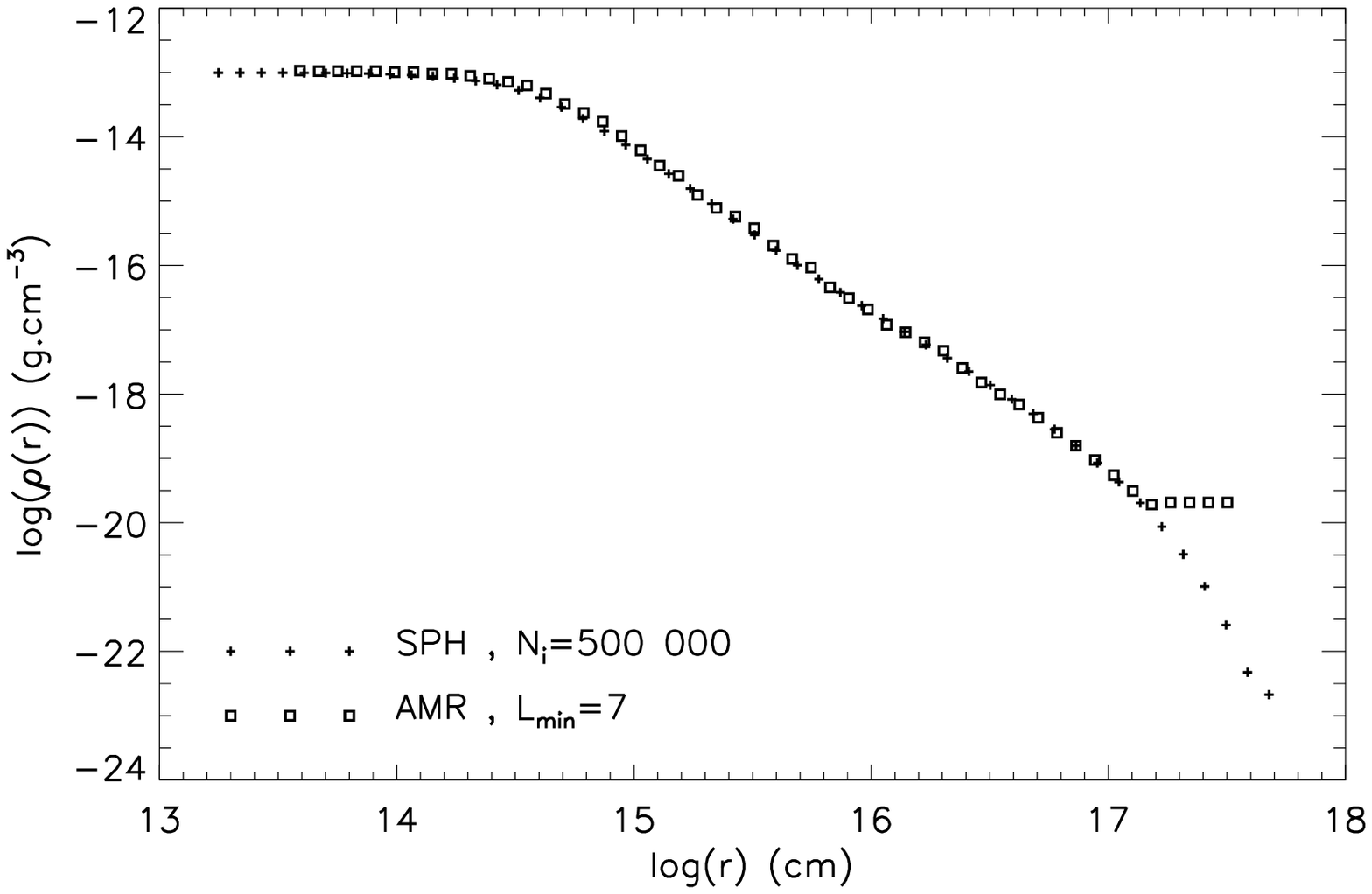}
  \includegraphics[width=8cm,height=6cm]{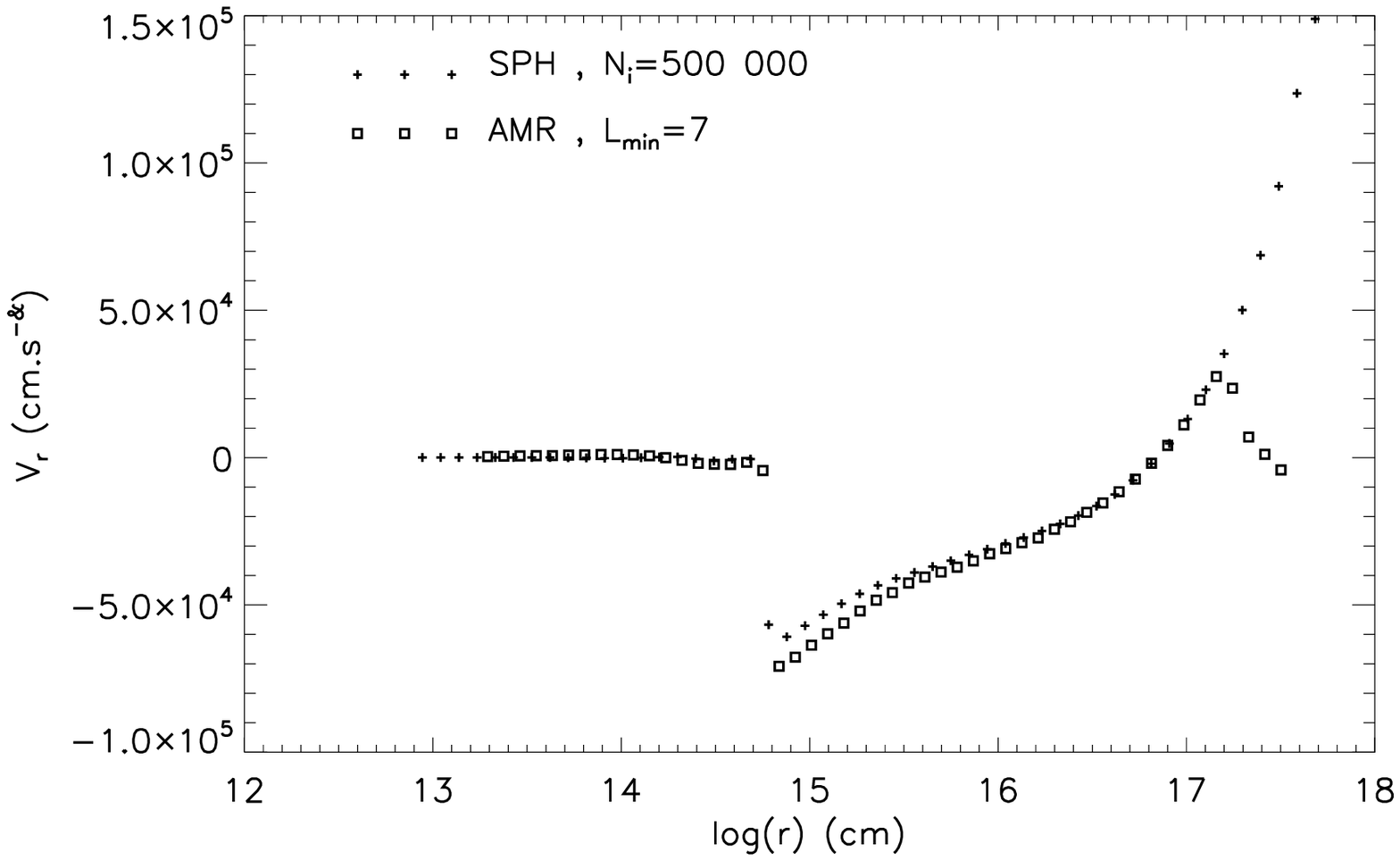}
  \caption{(a): \rm  Density profiles  at t$_0$ as  a function  of the
radius for  the case with $\beta=0.01$  and for the  two most resolved
simulations,  namely $\ell_\mathrm{min}=7$, $N_\mathrm{J}=10$  for the
AMR  (squares) and  $N_\mathrm{p}=5\times10^5$,  $N_\mathrm{N}=50$ for
the SPH (crosses).   Fig.\ref{dens_accr_shock}(b): \rm illustration of
the  accretion  shock:  radial  velocity  profiles for  the  two  most
resolved AMR  and SPH simulations in  the $xy$-plane at  $\sim $t$_0 +
1.2$ kyr. }
\label{dens_accr_shock}
\end{figure}

\begin{figure}[t]
  \centering
  \includegraphics[width=8cm,height=5cm]{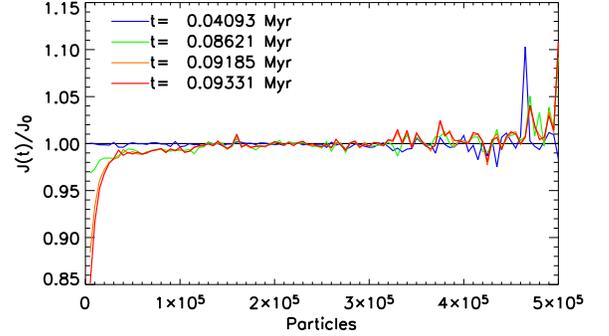}
   \caption{Ratio between  the angular momentum  $\vec{J}(t)$ over the
 initial angular  momentum $\vec{J_0}$ at different times  for the SPH
 simulation with $N_\mathrm{p}=5\times10^5$ and $N_\mathrm{N}=50$. The
 ratio is plotted as a function of the number of particles, ordered in
 decreasing  density. The  value of  the ratio  is a  mean  value over
 7\,500 particles. The red curve corresponds to results at t$_0$.}
\label{mom_sph}
\end{figure}

\subsection{Free-fall time}

The first step  is to compare the calculations  collapse time when the
initial  sphere is not  rotating (i.e.   $\beta=0$).  Note  that since
$\alpha$ is large and since we  use a barotropic equation of state, we
expect to find  a value larger than the free-fall  time.  Then, we use
as   a  reference   time   t$_0\ga  ($   t$_\mathrm{ff})$  for   which
$\rho_\mathrm{max}=\rho_c$.  Table \ref{sum_ff}  gives  collapse times
t$_0$  as  a  function  of  resolution  parameters  for  AMR  and  SPH
calculations.  AMR  calculations have been  run with $N_\mathrm{J}=10$
and   $\ell_\mathrm{min}=5,6$    and   7,   SPH    calculations   with
$N_\mathrm{p}$  ranging from $5\times  10^3$ to  $2\times 10^5$  and a
number   of   neighbors   $N_\mathrm{N}=50$.    We  note   that   with
$N_\mathrm{p}=5\times10^3$,     we      do     not     satisfy     the
\cite{bate_burkert-1997}  criterion, but  the mass  of  our resolution
element, i.e.  the sphere  containing the $N_\mathrm{N}$ neighbors, is
smaller that the critical Jeans mass.

With  increasing  numerical   resolution,  the  numerical  time  t$_0$
decreases and seems  to converge toward a value  slightly greater than
the free-fall  time. Time t$_0$ changes  by less than  5\% between AMR
calculations  with $\ell_\mathrm{min}=6$  and 7  and  SPH calculations
with $N_\mathrm{p}=5\times 10^4$  and $2\times 10^5$.  Dynamical times
t$_0$ in  SPH calculations are closer  to the free-fall  time than the
AMR ones. This is partly due to the higher initial resolution in SPH.

\subsubsection{Collapse and accretion shock with rotation $\beta=0.01$}

The gas  sphere is now in  solid rotation around the  $z$-axis. We set
$\beta=0.01$,  corresponding to  an  orbital time  t$_\mathrm{rot}=2.8
\times 10^{3}$ kyr.   Table \ref{sum_homorot} summarizes the different
SPH and  AMR calculations run for  this case.  In  order to illustrate
the   core  resolution,   we  give   the   quantity  $N_\mathrm{core}$
representing the number of cells/particles with density $\rho> 1\times
10^{-15}$ g.cm$^{-3}$  at t$_0$.  AMR simulations  have been performed
with different  minimum refinement levels  $\ell_\mathrm{min}$ ranging
from 5 to  7 and a refinement criterion  $N_\mathrm{J}$ ranging from 4
to  10.  As  expected, at  a constant $N_\mathrm{J}$, the  various AMR
calculations show  a convergence.  The SPH simulations  were performed
with  a constant  number of  neighbors $N_\mathrm{N}=50$  and  a total
number  of  particles  $N_\mathrm{p}$  ranging from  $5\times10^3$  to
$5\times10^5$.   A consequence  of the  deteriorating  resolution with
increasing  density   in  standard  SPH,  as   mentioned  earlier,  is
illustrated by the fact that,  for equivalent initial condition, it is
easier to get a better core resolution with AMR.

In  Fig.\ref{dens_accr_shock}(a) we  show  density  profiles as  a
function of  the radius in  the equatorial plane for  AMR calculations
with $\ell_\mathrm{min}=7$ and $N_\mathrm{J}=10$  and for the SPH with
$N_\mathrm{p}=5\times10^5$   and   $N_\mathrm{N}=50$.    The   density
profiles are very similar, indicating good convergence between the two
methods.  The  behaviour differs at relatively high  radius because of
the external gas  in the AMR method.  In  the present simulations, the
dynamical  time to reach  $\rho_\mathrm{c}$ is  increased by  $\sim 5$
kyr,   because  of  the   rotational  support.    As  seen   in  Table
\ref{sum_homorot},  when one  increases the  resolution, one  seems to
converge toward this time.

\begin{figure*}[htb]
  \centering
  \includegraphics[width=16cm,height=6cm]{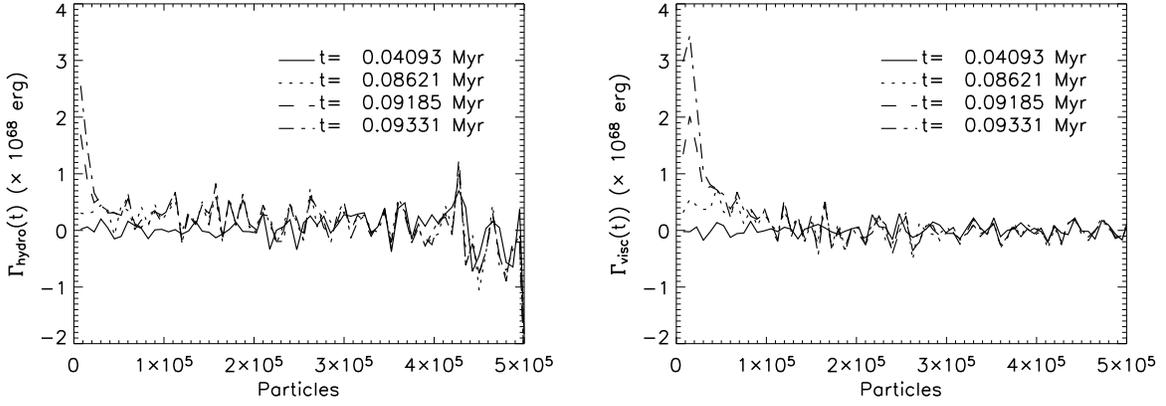}
  \caption{\it{Left plot:} \rm  Cumulated hydrodynamical torque on the
rotational  axis for  SPH  calculations with  $N_\mathrm{p}= 5  \times
10^{5}$  and   $N_\mathrm{N}=50$  and at the  same  times  as  in  Fig.
\ref{mom_sph}.    \it{Right  plot:}  \rm   Cumulated  torque   on  the
rotational axis due to the  standard artificial viscosity for the same
calculations and  times. Particles  are ordered in  decreasing density
and  the torques  are averaged  over 7\,500  particles.  At  t$_0$ the
cumulated torques are significant for the densest particles.  }
\label{couples}
\end{figure*}

When  the core  becomes adiabatic,  the angular  momentum conservation
induces the formation of an  accretion disk around the central object.
The centrifugal  force becomes comparable to the  gravitational one on
the equatorial plane, slowing  down the collapse. The outer collapsing
gas, which has a supersonic infalling speed, meets suddenly the static
gas  of the  core, creating  an accretion  shock.  This  shock  can be
clearly  seen  in   Fig.  \ref{dens_accr_shock}(b)  where  the  radial
velocity component  averaged over  the equatorial plane  is displayed.
The accretion shock is described slightly more accurately with the AMR
method.   The  SPH curve  is  smoother before  the  shock  due to  the
artificial  viscosity scheme.   The  slope before  the shock  strongly
depends  on the hydrodynamical  solver so  the results  illustrate the
difference between the hydrodynamical methods used in our two codes.

\subsection{{\it Theoretical} local angular momentum}

We now  investigate the issue  of angular momentum  conservation. Note
that  both SPH  and AMR  equations ensure  conservation of  the linear
momentum.   Considering  our  axisymmetric  model,  without  azimuthal
perturbation,  we  can  easily  investigate the  effect  of  numerical
resolution  on  angular  momentum  conservation.   The  local  angular
momentum should be well conserved, until azimuthal symmetry is broken.
The  loss of  local  angular momentum  in  our model  is  only due  to
unphysical transport inherent to the numerical methods used in the two
codes.   Thanks to  its  Lagrangian properties,  the SPH  calculations
gives access to the angular momentum that each particle has initially,
i.e.  the angular momentum that  particle should have if the numerical
scheme  was conserving  it  exactly.  Having  access  to the  particle
initial  angular momentum,  the  loss of  angular  momentum is  easily
calculated.  The azimuthal velocity  component, directly linked to the
angular momentum $\vec{J}$, is given by

\begin{equation}
v_\theta = \frac{xv_y - yv_x}{r} ,
\end{equation}
where $r$ is the distance from the rotation axis 
\begin{equation}
r = \sqrt{x^2 + y^2},
\end{equation}
and the angular momentum 
\begin{equation}
\vec{J} = xv_y - yv_x.
\end{equation}

The  angular momentum  conservation along  the $z$-axis  allows  us to
write
\begin{equation}
\vec{J}= xv_y - yv_x = x_0v_{y,0} - y_0v_{x,0} = \vec{J}_0.
\end{equation}

Thus, the  theoretical angular velocity of  a particle at  time $t$ is
determined  by   the  ratio  between  its   initial  angular  momentum
$\vec{J}_0=(x_0v_{y,0} - y_0v_{x,0})$ and its actual radius:
\begin{equation}
v_{\theta,\mathrm{th}} = \frac{\vec{J}_0}{r}.
\label{momth}
\end{equation}

With  Eq.   (\ref{momth}),  we  can compare  the  theoretical  angular
velocity component to  the numerical ones, and in  particular with the
AMR results  for which we do  not have access to  a theoretical value.
Since SPH and  AMR density profiles are almost  identical at t$_0$, we
suppose  that  the  previous  mapping  giving  $\vec{J}_0(r_0)$  as  a
function  of the radius  in the  SPH runs  is also  valid for  the AMR
calculations.   Note  also that  the  method  cannot  account for  the
displacement  of  the particles  which  would  arise  by changing  the
angular momentum.

\subsubsection{Azimuthal velocity component}\label{mom_ang}

Figure  \ref{mom_sph} gives  the  ratio between  the angular  momentum
$\vec{J}(t)$   and   $\vec{J}_0$   for   the   SPH   simulation   with
$N_\mathrm{p}=5\times 10^5$  and $N_\mathrm{N}=50$. The  particles are
ordered in  decreasing density and  the ratio is averaged  over $7500$
particles. A  first interesting result is that  denser particles loose
more angular momentum.  At t$_\mathrm{ff}$, denser particles have lost
3\% of their initial  angular momentum. In less resolved calculations,
i.e. $N_\mathrm{p}  = 5\times 10^3$ particles, the  effect is stronger
and the  densest ones  loose more than  10\% of their  initial angular
momentum  in  a   free-fall  time  t$_\mathrm{ff}$.   This  percentage
slightly  decreases  when increasing  the  number  of particles.   The
reason is that  the denser the particle the  larger the viscous torque
(see below) and thus the  larger the angular momentum transport.  This
numerical transport  is amplified when  the core is close  to becoming
adiabatic.

Figure  \ref{couples} shows  the  cumulated hydrodynamical  (left-hand
side) and viscous (right-hand side) torques on the rotational axis for
the same SPH calculations at  the same times as in Fig. \ref{mom_sph}.
In principle,  there should not be  any torque on  the rotational axis
because  of  this  axisymmetric  model.   The  cumulated  torques  are
computed and  summed at  each timestep for  each particle.   The value
plotted is an average over  7\,500 particles and particles are ordered
in decreasing  density.  Denser  particles have the  largest cumulated
hydrodynamical and viscous torques.  The friction forces corresponding
to the  hydrodynamical torque are stronger for  these particles, which
is due to the strong differential velocity.


\begin{figure*}[htb]
  \centering
  \includegraphics[width=8cm,height=6cm]{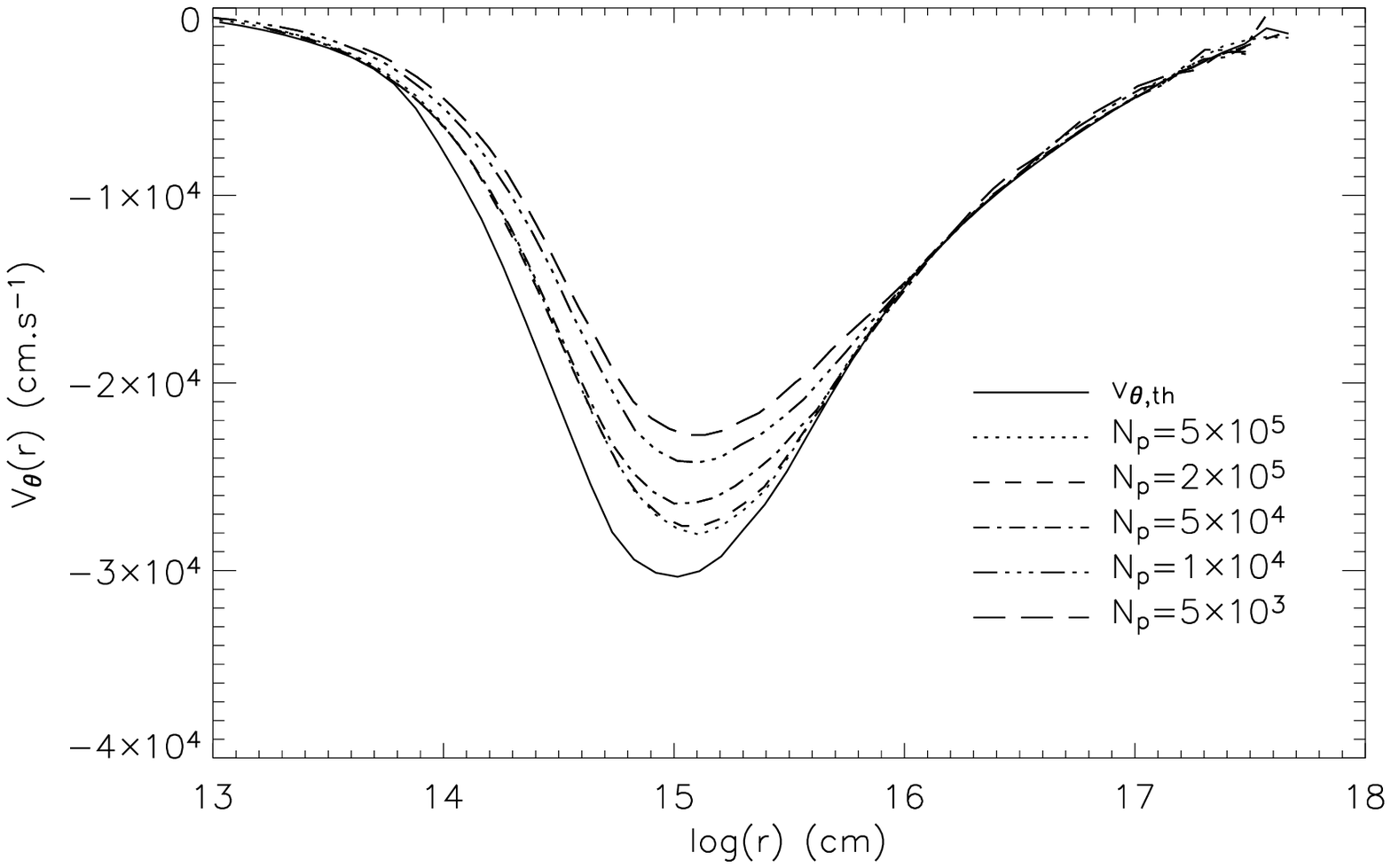}
  \includegraphics[width=8cm,height=6cm]{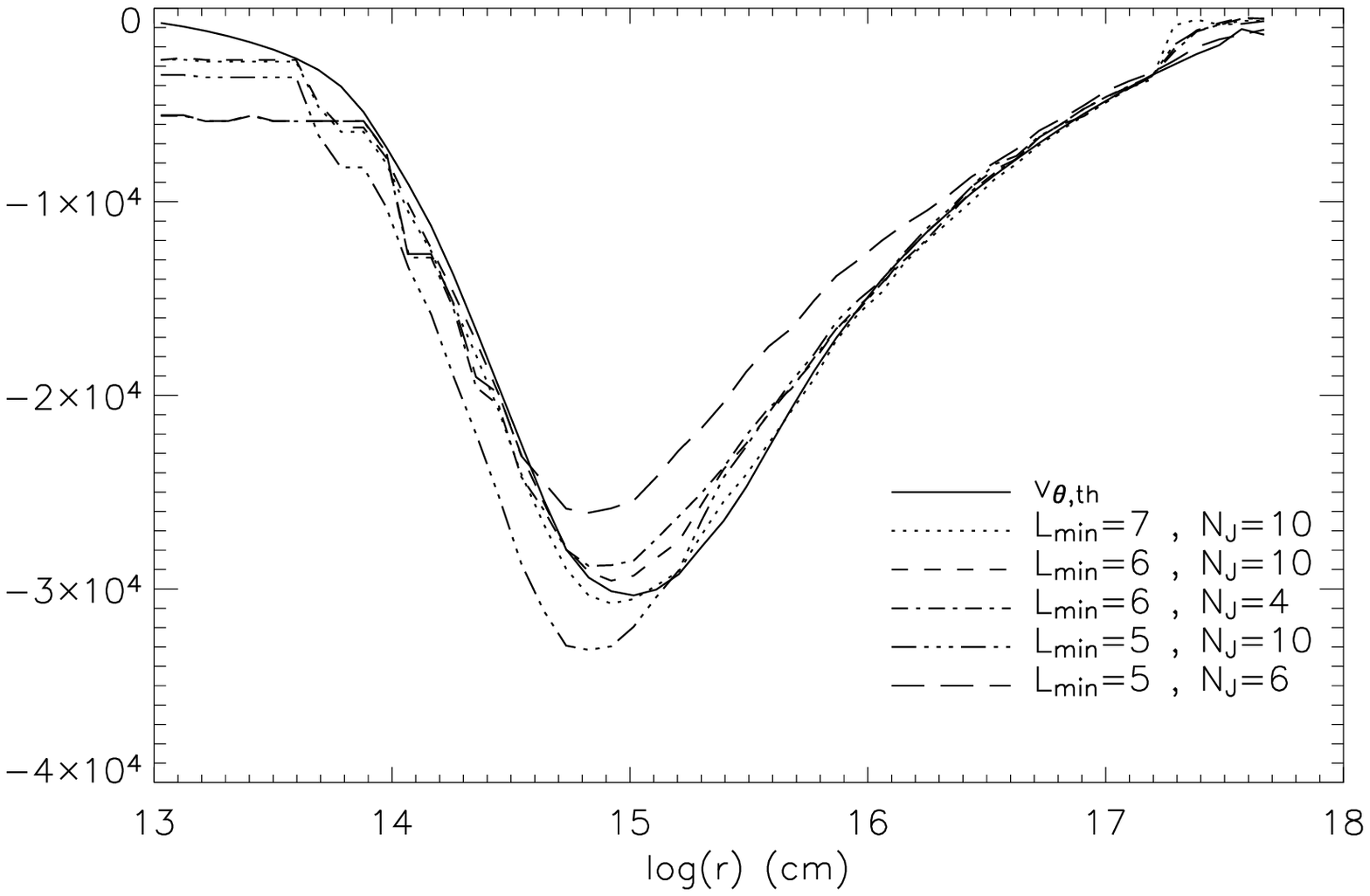}
  \caption{Azimuthal velocity at t$_0$ as  a function of the radius on
the equatorial  plane for SPH  (left) and AMR (right)  calculations at
corresponding t$_0$.   The left-hand plot  (Fig. \ref{homorot}a) shows
SPH  results with various  $N_\mathrm{p}$ and  $N_\mathrm{N}=50$.  The
solid line represents  the theoretical azimuthal velocity interpolated
at t$_0$  and is denoted as  $v_{\theta,\mathrm{th}}$.  The right-hand
plot (Fig.   \ref{homorot}b) shows AMR  results with $N_\mathrm{J}=10$
and  $\ell_\mathrm{min}=5$,  6   and  7.   The  theoretical  azimuthal
velocity is plotted also for easy comparison with the SPH results.}
\label{homorot}
\end{figure*}

\begin{table}[htb]

\caption{Summary of the different simulations (upper table: SPH, lower
table: AMR) performed to study angular momentum conservation.}
\centering

\begin{tabular}{c}

\begin{tabular}{ccccc}
\hline
\hline
$N_\mathrm{p}$    & $N_\mathrm{N}$ & $N_\mathrm{J} $ & $N_\mathrm{core}$& t$_0$ (kyr) \\
\hline
$5 \times 10^3$  & 50    & 1.86 & 225   & 115 \\
$1\times 10^4$   & 50    & 2.34 & 422   & 107 \\
$5\times 10^4$   & 50    & 4.   & 1 833 &  98 \\
$2\times 10^5$   & 50    & 6.35 & 7 055 &  95 \\
$5\times 10^5$   & 50    & 8.61 & 17 309&  93 \\
\hline
\end{tabular} 
\\
\\
\space
\begin{tabular}{cccccc}
\hline
\hline
 $\ell_\mathrm{min}$   & $N_\mathrm{i}$ & $N_\mathrm{J}$  & $N_\mathrm{core}$ & Tot. cells & t$_0$ (kyr)\\
\hline 
5   &   2 145    &  6 &  3 928 & $\sim 9.1 \times 10^4$ & 150 \\
5   &   2 145    & 10 & 30 752 & $\sim 1.6 \times 10^5$ & 116 \\
6   &  17 160    &  4 &  4 016 & $\sim 3.1 \times 10^5$ & 116 \\
6   &  17 160    & 10 & 28 800 & $\sim 3.7 \times 10^5$ & 109 \\
7   & 137 260    & 10 & 29 944 & $\sim 2.2 \times 10^6$ &  96 \\
\hline
\end{tabular}
\end{tabular}
\label{sum_homorot}
\end{table}

Figure \ref{homorot}a  displays the azimuthal velocity  component as a
function of the radius $r$  on the $xy$-plane for the SPH simulations.
The theoretical  azimuthal velocity  profile (solid line)  is obtained
following  the previous section.   It is  obvious that  low resolution
simulations are  not able to  conserve properly the  angular momentum.
With $5\times 10^3$ particles, we obtain counter-rotating particles at
the  center (not  illustrated in  Fig. \ref{homorot}a  because  of the
average  in the $xy$-plane  that smoothes  the profiles).   It appears
that a  minimum of  $5\times 10^4$ particles  is required  to maintain
angular momentum  loss within less than  $10 \%$, for the  case of the
present  study.   The  improvement  of angular  momentum  conservation
eventually saturates for large  numbers of particles.  We checked that
using a larger  number of neighbors does not  improve the conservation
of local angular momentum.

Figure \ref{homorot}b shows results obtained with the RAMSES code. AMR
curves  are plotted  and compared  with the  theoretical  one obtained
previously  for SPH.  The  simulations with  $\ell_\mathrm{min}=6$ and
$\ell_\mathrm{min}=7$ are close to the theoretical curve.  In both AMR
simulations  with $N_\mathrm{J}=10$, dense  core resolution  is higher
than with  SPH.  For  an initial resolution  of $\ell_\mathrm{min}$=5,
the AMR  scheme does produce some  angular momentum lag.   This can be
due to the fact that with a poor initial resolution, the interpolation
of the  gravitational potential tends to  convert gravitational energy
into  rotational energy.  The outer  gas  does not  alter the  angular
momentum  conservation  for  AMR  calculations  because  of  its  tiny
density.

\begin{figure}[htb]
  \centering
  \includegraphics[width=8cm,height=6cm]{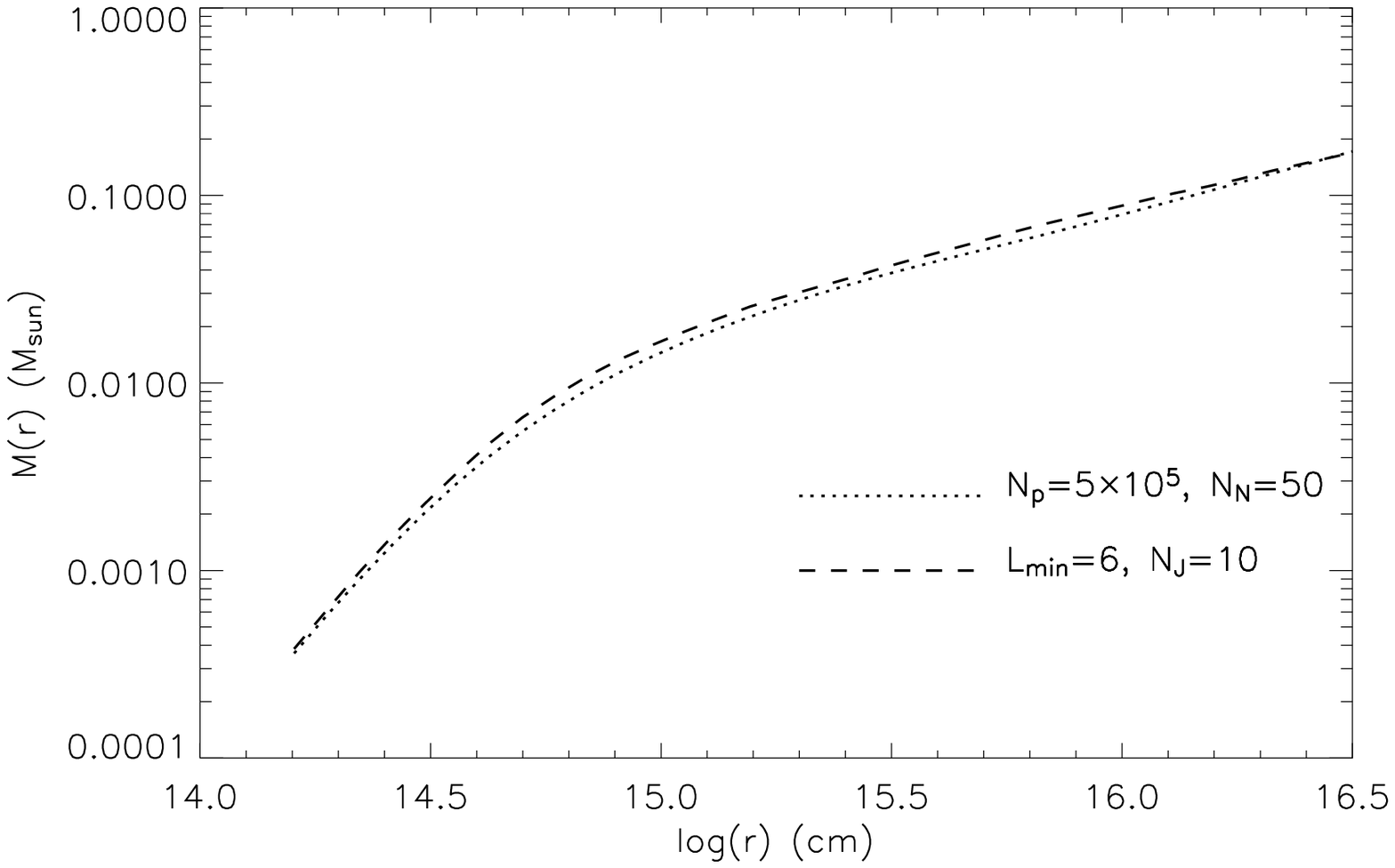}
  \includegraphics[width=8cm,height=6cm]{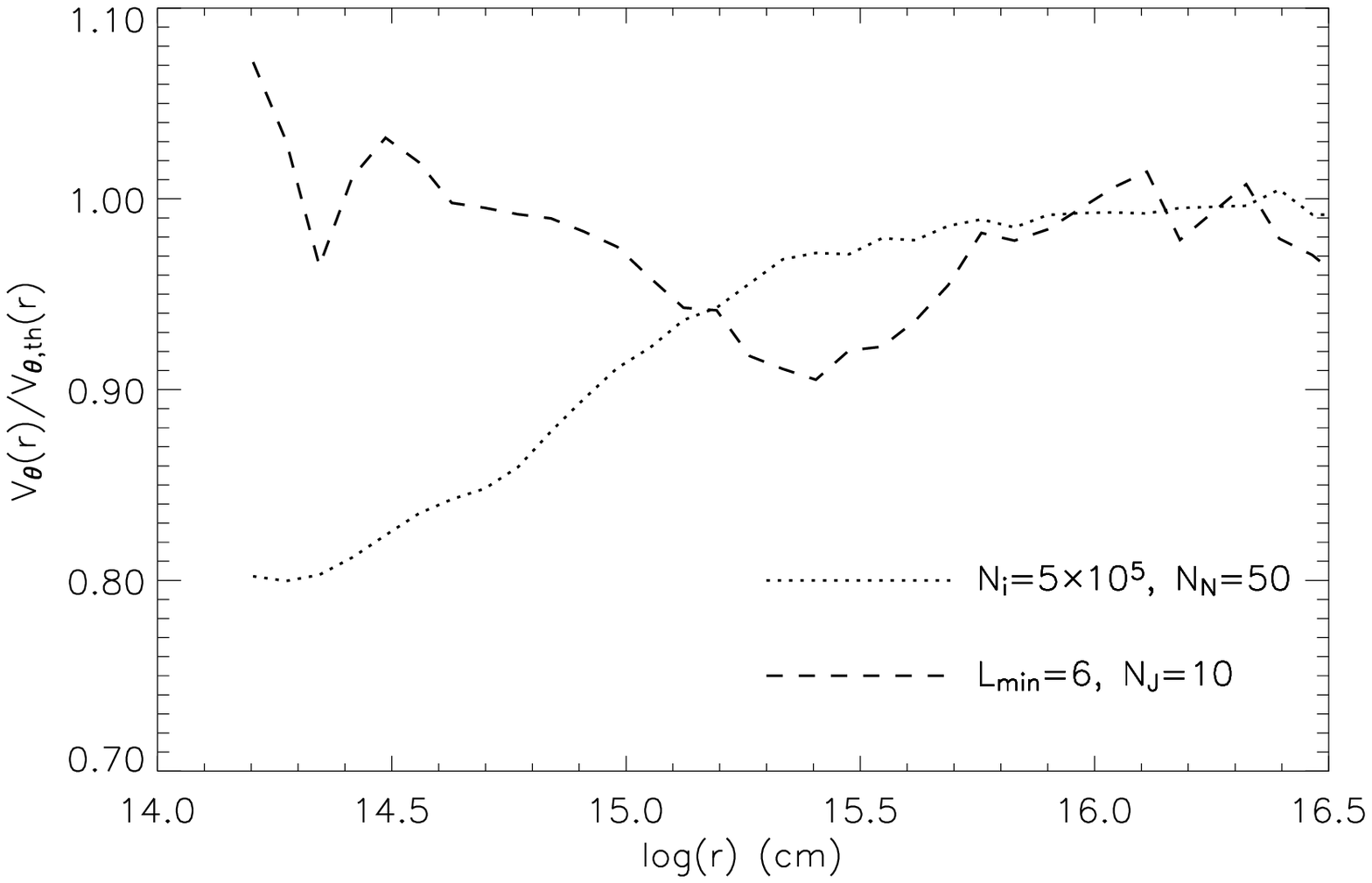}
  \caption{(a): Total integrated mass in the equatorial plane at t$_0$
as   a   function   of   the   radius  for   AMR   calculations   with
$\ell_\mathrm{min}=6$,  $N_\mathrm{J}=10$  (dashed-line)  and for  SPH
calculations   with  $N_\mathrm{p}=5\times   10^5$,  $N_\mathrm{N}=50$
(dotted line).  \rm(Fig.\ref{mom_trans}b): Ratio between numerical and
theoretical azimuthal velocities at t$_0$  as a function of the radius
in the equatorial plane for the same calculations.}
\label{mom_trans}
\end{figure}

In Fig.  \ref{mom_trans}a,b, we plot the integrated mass and the ratio
of numerical  over theoretical  angular momentum for  AMR calculations
with $\ell_\mathrm{min}=6$, $N_\mathrm{J}=10$ and for SPH calculations
with     $N_\mathrm{p}=5\times    10^5$,     $N_\mathrm{N}=50$.     In
Fig. \ref{mom_trans}a,  most of the  mass of the forming  disk remains
within a radius $\sim 1\times 10^{16}$ cm, i.e. where AMR and SPH have
an opposite behaviour.  In Fig.  \ref{mom_trans}b, angular momentum is
clearly transported  to the outer regions  with SPH, whereas  it has a
slight trend to be transported to the inner regions with the AMR.  The
overall angular  momentum is well conserved in  both calculations, but
local  properties seem  to  be affected  by  initial resolution.   SPH
resolution is much better at  the beginning of the calculations, which
enables  SPH  to  properly  conserve  local angular  momentum  at  $r>
1.4\times 10^{15}$  cm at  t$_0$ while low  initial resolution  of the
sphere in AMR  induces a worse conservation.  This  effect reverses at
lower radius where  AMR can reach smaller scales  contrary to standard
SPH.

The density profiles obtained with  the two methods converge towards a
similar  solution  (c.f.   Fig.\ref{dens_accr_shock}a)  but  a  closer
analysis of the velocity profiles shows discrepancies. We can say that
angular  velocity profiles  indicate  that local  angular momentum  is
better conserved  with the AMR than  with the SPH method  for the code
implementation that  we used.  Both can be  improved using appropriate
methods (see appendix \ref{app_visc} and \ref{note_diff}).

\section{Fragmentation}
\subsection{Model}

Prestellar core  fragmentation is a highly non-linear  process.  A key
issue is to  understand to what extent the  fragmentation which occurs
in  numerical simulations is  influenced by  the numerical  scheme and
resolution.

\begin{figure*}[htb]
  \centering
  \includegraphics[width=17.cm,height=17cm]{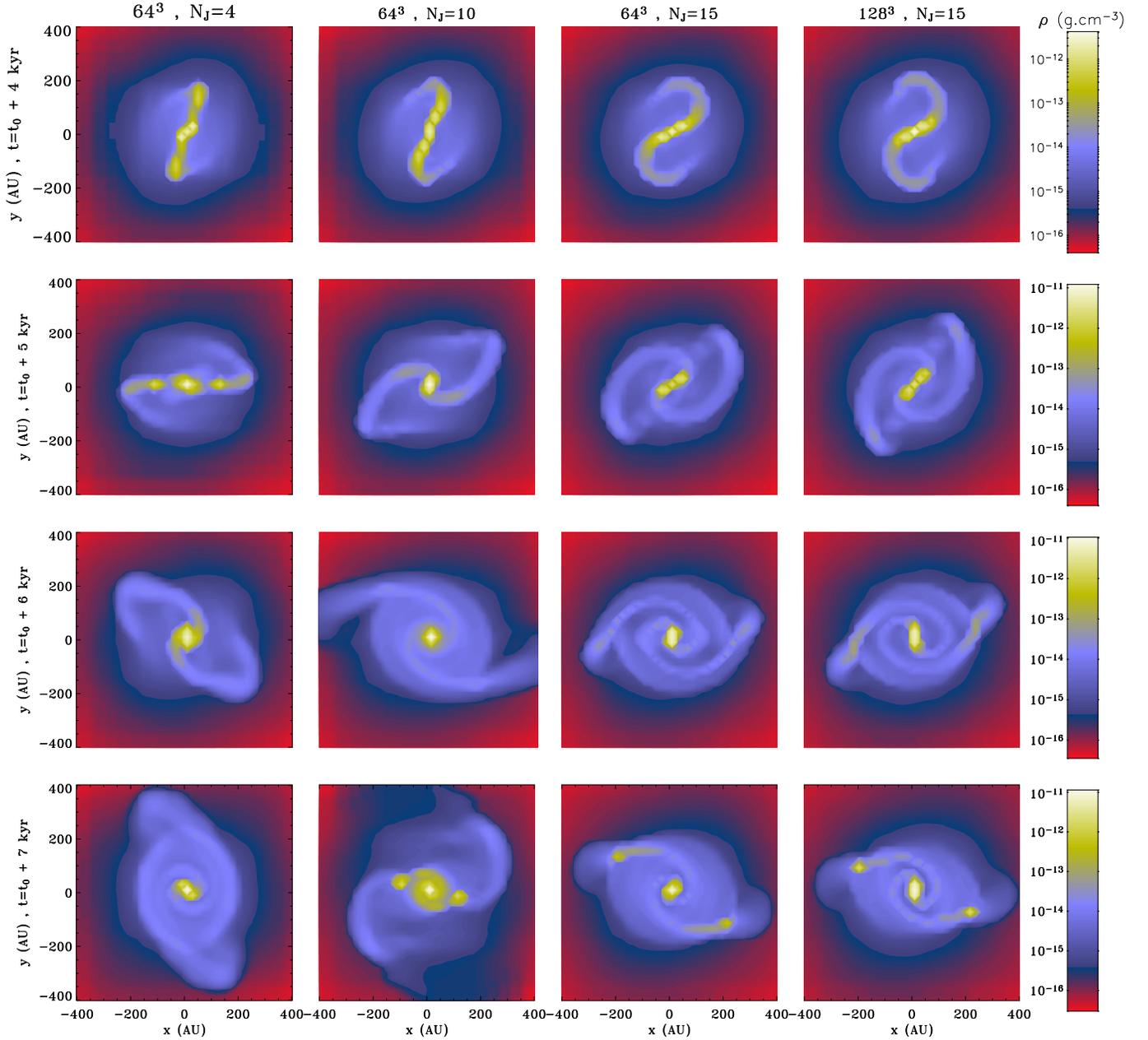}
  \caption{AMR  calculations density  maps on  the $xy$-plane  for the
case $\alpha=0.50$,  $\beta=0.04$. From top to  bottom, four different
times are showed:  t=t$_0 + 4$ kyr,  t=t$_0 + 5$ kyr, t=t$_0  + 6$ kyr
and t=t$_0 +  7$ kyr.  The AMR calculations  have been performed with,
from    left    to    right   columns,    $\ell_\mathrm{min}=6$    and
$N_\mathrm{J}=4$,     $N_\mathrm{J}=10$,     $N_\mathrm{J}=15$     and
$\ell_\mathrm{min}=7$ and $N_\mathrm{J}=15$.}
\label{amr_a050}
\end{figure*}
  
To  study  dense  core  fragmentation,  we choose  the  same  previous
spherical  model   and  impose  a   $\mathrm{m}=2$  azimuthal  density
perturbation:
\begin{equation}
\rho(\theta)=\rho_0[1 + A \, \mathrm{cos} (\mathrm{m}\theta)] ,
\end{equation}
where  $\rho_0$  is the  mean  sphere  density,  $A$ the  perturbation
amplitude and $\theta$ the azimuthal angle in cylindrical coordinates.
\newline

The initial conditions are easy to implement for the AMR calculations.
The SPH  sinusoidal density perturbation  is imposed by  adjusting the
unperturbed $\theta$-coordinate of each  particle to a perturbed value
$\theta^*$ given by:
\begin{equation}
\theta = \theta^* +  \frac{A \, \mathrm{sin}(\mathrm{m}\theta^*)}{\mathrm{m}}.
\end{equation}

We compare  simulations for  three different thermal  supports, namely
$\alpha = 0.35$, 0.5 and  0.65, with a fixed rotational support $\beta
= 0.04$.  In the following,  the convention is to call {\it fragments}
the clumps  where the gas  density satisfies $\rho >  1\times 10^{-12}$
g.cm$^{-3}$.

\subsection{Results for a critical case: $\alpha=0.5 , \beta=0.04$ and A=0.1}

\begin{table}
\caption{Summary   of   the   different  simulations   performed   with
$\alpha=0.5$   and  $\beta=0.04$.  $N_\mathrm{core}$   corresponds  to
particles/cells  whose  density  satisfies  $\rho >  1\times  10^{-15}$
g.cm$^{-3}$   at   t=t$_0  +   4$   kyr.   The   upper  table   (Table
\ref{sum_a050}a) gives  a summary of  the SPH calculations.  The lower
Table (Table \ref{sum_a050}b) shows  a summary of AMR calculations. }
 \centering
\begin{tabular}{c}
\begin{tabular}{cccc}
\hline
\hline
$N_\mathrm{p}$    & $N_\mathrm{N}$  & $N_\mathrm{core}$ & t$_0$ (kyr)\\
\hline
$5\times 10^4$ & 50     &  6 997 & 64 \\
$2\times 10^5$ & 50     & 28 516 & 63 \\
$5\times 10^5$ & 30     & 72 417 & 63 \\
$5\times 10^5$ & 50     & 71 804 & 62 \\
$5\times 10^5$ & 100    & 76 390 & 63 \\
$5\times 10^5$ & 200    & 71 940 & 64 \\
\hline
\end{tabular}
\\
\\
\begin{tabular}{ccccc}
\hline
\hline
 $\ell_\mathrm{min}$ & $N_\mathrm{J}$  & $N_\mathrm{core}$ & Tot. cells & t$_0$ (kyr)\\
\hline 
6   &  4 & 48 680 & $\sim 4.4  \times 10^5$ & 66 \\
6   & 10 & 156 588 & $\sim 6.2 \times 10^5$ & 65 \\
6   & 15 & 263 304 & $\sim 7.9 \times 10^5$ & 67 \\
7   & 10 & 117 108 & $\sim 2.4 \times 10^6$ & 65 \\
7   & 15 & 305 896 & $\sim 2.7 \times 10^6$ & 65 \\
\hline
\end{tabular}
\end{tabular} 
\label{sum_a050}
\end{table}

This  subsection is devoted  to the  exploration of  various numerical
parameters. First,  we study  the effect of  varying the  initial grid
resolution $\ell_\mathrm{min}$ and the  number of cells within a Jeans
length $N_\mathrm{J}$ for AMR  calculations.  Then, we present our SPH
calculations with  various number  of neighbors $N_\mathrm{N}$  and of
particles $N_\mathrm{p}$.   For this set of  calculations, the initial
parameters    are:     $\rho_0=1.35\times    10^{-18}$    g.cm$^{-3}$,
$R_0=7.07\times    10^{16}$    cm,   $\Omega_0=2.12\times    10^{-13}$
rad.s$^{-1}$  and t$_\mathrm{ff}=57$  kyr.   The initial  perturbation
amplitude is $A=0.1$.

Table \ref{sum_a050} summarizes the calculations we performed for this
case.   Informations about  the core  resolution  (i.e.  $\rho>1\times
10^{-15}$  g.cm$^{-3}$) and  the total  number of  cells are  given at
t=t$_0 + 4$ kyr.  The dynamical  times to reach the collapse are quite
similar, within less than 2\%.  Synchronizing calculations at t$_0$ is
then  well justified.   In the  following sections,  we  consider core
evolutions over a few thousand years ($\sim 10 \%$ of t$_0$).

\subsubsection{Detailed study of the effect of $N_\mathrm{J}$ and $\ell_{min}$ for AMR calculations}

In Fig. \ref{amr_a050},  we show density maps on  the equatorial plane
at four  different timesteps for  AMR calculations with, from  left to
right,  $\ell_\mathrm{min}=6$  and  $N_\mathrm{J}=4$,  10 and  15  and
$\ell_\mathrm{min}=7$    (giving   128$^3$   cells    initially)   and
$N_\mathrm{J}=10$.  Maps  are given, from top to  bottom, at t=t$_0+4$
kyr, t=t$_0+5$  kyr, t=t$_0+6$  kyr, and t=t$_0+7$  kyr.  As  shown in
Fig.  \ref{amr_a050}, fulfilling the Truelove condition ($N_\mathrm{J}
> 4$)  does not  guarantee an  accurate fragmentation  timescale.  The
calculations with  $N_\mathrm{J}=4$ fragments at  t$_0 + 5$  kyr while
other  calculations do  not  fragment  until t=t$_0  +  7$ kyr.   This
suggests    that    calculation    with   $\ell_\mathrm{min}=6$    and
$N_\mathrm{J}=4$  suffers  from  inaccurate  fragmentation,  but  this
latter is inhibited  when fragments fall on the  central object before
t$_0+6$ kyr.  The core will  eventually refragment but not at the same
time as  the other calculations  (i.e.  t $>>$  t$_0 + 7$  kyr).  With
increasing $N_\mathrm{J}$,  we converge  to a fragmented  pattern with
one central object and two satellites.

Another  aspect to  be considered  quite  carefully is  the choice  of
$\ell_\mathrm{min}$, i.e. the initial  description of the sphere.  The
two calculations with  $N_\mathrm{J}=15$ and $\ell_\mathrm{min}=6$ and
$\ell_\mathrm{min}=7$  are  very  similar, suggesting  that  numerical
convergence  has been  achieved. Even  though small  differences still
appear  in  the  detailed  structures.   The  satellites  formed  with
$\ell_\mathrm{min}=7$  are  more  structured  and compact  than  those
formed with  the initial resolution  $\ell_\mathrm{min}=6$.  According
to these calculations, fragmentation into two identical satellites and
a central object should occur around t=t$_0+7$ kyr.

\subsubsection{Detailed study of the effect of $N_\mathrm{p}$ and $N_\mathrm{N}$ for SPH calculations}

\begin{figure*}[htb]
  \centering
  \includegraphics[width=17.cm,height=17cm]{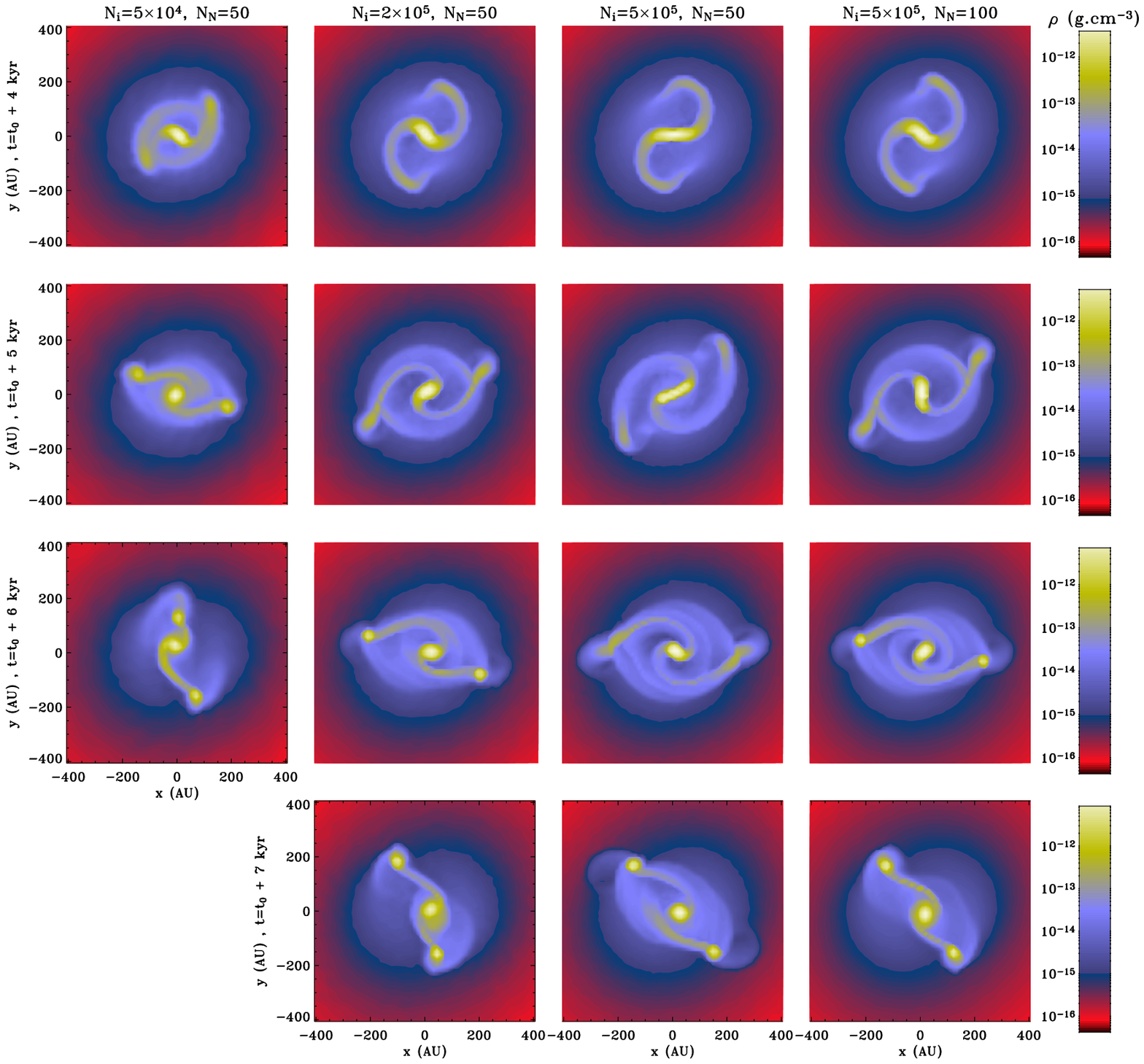}
  \caption{SPH calculations density maps in the $xy$-plane for the case
$\alpha=0.50$, $\beta=0.04$. From top  to bottom, four different times
are showed:  t=t$_0 + 4$  kyr, t=t$_0  + 5$ kyr,  t=t$_0 + 6$  kyr and
t=t$_0 + 7$ kyr.  The calculations have been performed with, from left
to   right,   $N_\mathrm{N}=50$   and   $N_\mathrm{p}=5\times   10^4$,
$N_\mathrm{p}=2\times   10^5$  and  $N_\mathrm{p}=5\times   10^5$  and
$N_\mathrm{N}=100$ and $N_\mathrm{p}=5\times 10^5$.}
\label{sph_a050}
\end{figure*}

We   performed   a  series   of   SPH   calculations  with   different
$N_\mathrm{p}$, but with a constant $N_\mathrm{N}=50$, and then with a
larger $N_\mathrm{N}$ for one value of $N_\mathrm{p}$.

Figure \ref{sph_a050}  shows density maps on the  equatorial plane for
these calculations  at four different  timesteps, namely, from  top to
bottom: t=t$_0 + 4$ kyr, t=t$_0 + 5$ kyr, t=t$_0 + 6$ kyr and t=t$_0 +
7$  kyr.   The   calculations  with  $N_\mathrm{p}=2\times  10^5$  and
$N_\mathrm{p}=5\times 10^5$ show good agreement at least until t$_0$ +
6  kyr  and differ  noticeably  from  the  less resolved  calculations
($N_\mathrm{N}=50$, $N_\mathrm{p}=5\times  10^4$) where fragmenatation
occurs  earlier.   The dense  core  fragments  in  any case,  but  its
fragmentation    is     delayed    when    $N_\mathrm{p}$    increases
\citep[e.g.][]{Nelson_2006}.  Early fragmentation  is here clearly due
to  a lack  of resolution.   As  shown in  section \ref{mom_ang},  the
conservation of  local angular momentum is bad  when $N_\mathrm{p}$ is
low and leads to very inaccurate collapse and fragmentation timescales
of the cloud.  Once symmetry is  broken, it is useless to continue the
simulations, since calculations would obviously diverge. This symmetry
breaking occurs earlier  in the SPH calculations first  because of the
numerical noise inherent to  the relaxed, and random, initial particle
distributions, and also  because of the lower resolution  of the Jeans
length in the disk.

The other fundamental  parameter in SPH calculations is  the number of
neighbors determining  the kernel size.  Increasing the  number of SPH
particles increases the resolution  but also introduce numerical noise
at  smaller  scales.  The  natural  way to  reduce  this  noise is  to
increase  the smoothing  kernel length  by  increasing $N_\mathrm{N}$.
The  effects  of  varying  $N_\mathrm{N}$ have  been  investigated  by
\cite{Lombardi_1999}    and    \cite{rasio-1999}.    In    particular,
\cite{rasio-1999} derived the following results:
\begin{itemize}
\item  higher  accuracy  is   reached  when  both  $N_\mathrm{p}$  and
$N_\mathrm{N}$  are increased,  with $N_\mathrm{p}$  increasing faster
than  $N_\mathrm{N}$  so  that  the smoothing  length  decreases.  One
possible  scaling   \citep{Lombardi_1999}  is  $N_\mathrm{N}  \propto
N_\mathrm{p}^q$ with $0.2\leq q <1$,
\item    SPH   scheme    is    consistent   in    the   limit    where
$(N_\mathrm{N},N_\mathrm{p})  \rightarrow \infty$  and  $h \rightarrow
0$,
\item  convergence (e.g. the  number of  timesteps) is  accelerated by
increasing the smoothness of the kernel.
\end{itemize}  

The  usual   number  of  neighbors   in  previous  studies   is  about
50. However,  no  study  has   really  explored  the  role  played  by
$N_\mathrm{N}$ in the context  of star formation.  Hence, we performed
calculations with  a constant  $N_\mathrm{p}$ and different  values of
$N_\mathrm{N}$.   This is  illustrated  in the  right  column in  Fig.
\ref{sph_a050}   where   we   report   maps   of   calculations   with
$N_\mathrm{p}=5\times  10^5$  and  $N_\mathrm{N}=100$.  The  core  has
already fragmented  at t$_0+6$  kyr whereas with  $N_\mathrm{N}=50$ it
fragments    later.     The    calculations   with    parameter    set
$N_\mathrm{p}=2\times 10^5$  and $N_\mathrm{N}=50$ is  very similar to
the  later   of  similar  ratio   $N_\mathrm{p}/N_\mathrm{N}$.   Other
calculations  with  various $N_\mathrm{N}$  are  reported in  Appendix
\ref{note_Nn}.  It appears clearly that the greater $N_\mathrm{N}$ the
earlier fragmentation occurs,  because increasing $N_\mathrm{N}$ for a
fixed $N_\mathrm{p}$ decreases the spatial resolution ($h$ increases).

\subsubsection{Comparison and convergence}\label{comparison_a050}

\begin{figure}[htb]
  \centering
    \includegraphics[width=8.cm,height=10.67cm]{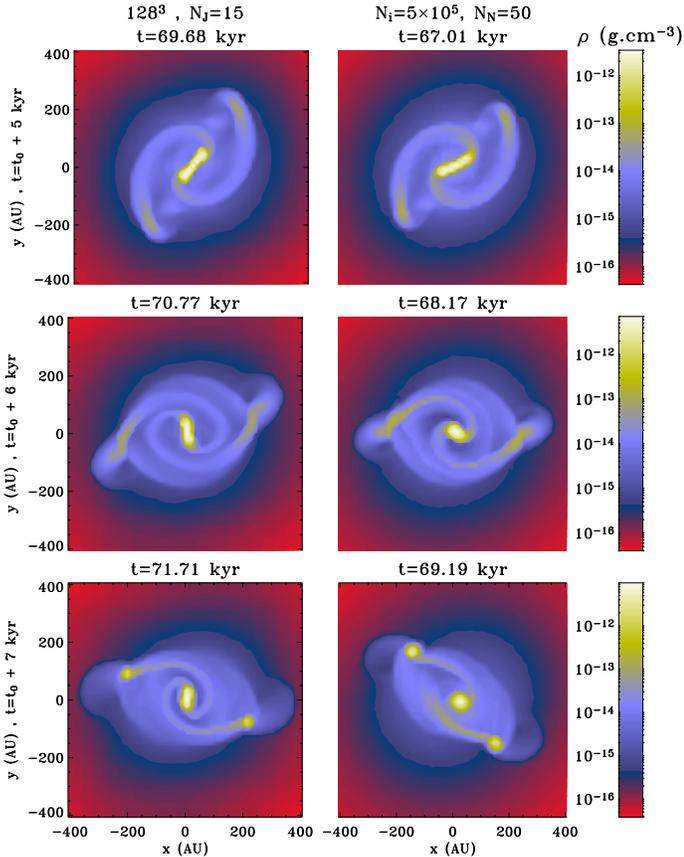}
  \caption{AMR and SPH calculations density maps in  the $xy$-plane at
three different  times for the case  $\alpha=0.50$, $\beta=0.04$.  The
times correspond to  t$_0 + 5$ kyr, t$_0  + 6$ kyr and t$_0  + 7$ kyr,
from  to bottom, respectively.   The AMR  calculations plotted  on the
left   column   have   been   run   with   $\ell_\mathrm{min}=7$   and
$N_\mathrm{J}=15$.   The right column  shows the  results for  the SPH
calculations with $N_\mathrm{p}=5\times 10^5$ and $N_\mathrm{N}=50$.}
\label{comp_a050}
\end{figure}

In  the previous  sections,  we  show that  AMR  and SPH  calculations
converged   separately.    We    now   cross-compare   the   converged
calculations.   Figure  \ref{comp_a050}  shows  density  maps  in  the
equatorial  plane for  the results  of two  amongst the  most resolved
calculations at  three timesteps, namely, from top  to bottom, t$_0$+5
kyr, t$_0$+6 kyr  and t$_0$+7 kyr. The left column  shows maps for AMR
calculations with  $\ell_\mathrm{min}=7$ and $N_\mathrm{J}=15$ whereas
the   right   column  displays   SPH   maps   for  calculations   with
$N_\mathrm{p}=5\times 10^5$  and $N_\mathrm{N}=50$.  We  display again
the results of Fig.\ref{amr_a050} and Fig.\ref{sph_a050} to illustrate
clearly the convergence.  Agreement  between the two methods for these
physical and  numerical parameters set  is striking for the  two first
timesteps.   The calculations  give  the same  fragmentation time  and
pattern, although  satellites and the  central object are  bigger with
the SPH.

\begin{figure}[htb]
  \centering
  \includegraphics{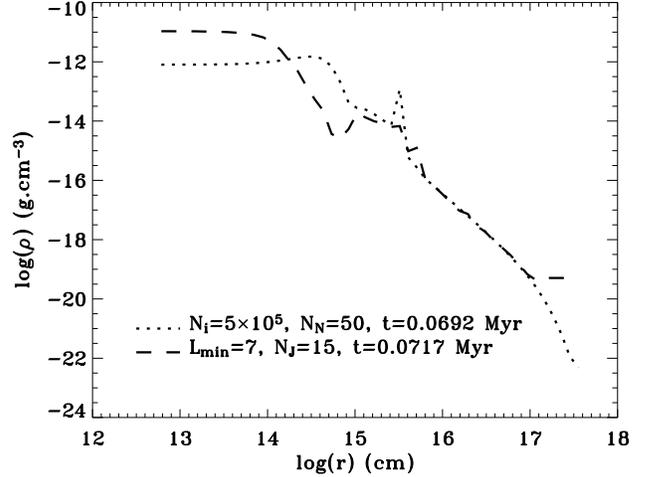}
  \caption{Density  profiles at  t$_0$+ 7  kyr  as a  function of  the
radius, averaged  on the equatorial  plane, for AMR  calculations with
$\ell_\mathrm{min}=7$  and $N_\mathrm{J}=15$  (dashed  line), and  SPH
calculations  with $N_\mathrm{p}=5\times10^{5}$  and $N_\mathrm{N}=50$
(dotted line).  }
\label{prof_dens_amr_a050}
\end{figure}

Figure  \ref{prof_dens_amr_a050}  shows  disk  density profiles  as  a
function of the  radius averaged in the equatorial  plane for the same
SPH and AMR calculations.   This plot complements Fig. \ref{comp_a050}
with   the    last   density   maps,   where    fragments   are   well
developed. Density profiles  show a peak at a  radius corresponding to
satellite positions in the map.  Satellite fragments are denser in the
SPH  calculations, and  the central  object is  less dense  and bigger
compared with the dense elongated shape obtained with the AMR.\\

Although there are some obvious  differences between the two methods ,
there  seems  to  be a  real  convergence  between  the two  types  of
calculations.   For the  specific case  under  study, we  find a  good
agreement between AMR calculations with $\ell_\mathrm{min}=7$ or 6 and
$N_\mathrm{J}=15$  and  SPH  calculations  with  $N_\mathrm{p}=5\times
10^5$  and $N_\mathrm{N}=50$.   However,  even for  the most  resolved
simulations, the  results between the  two methods diverge  after some
time  (i.e. t$_0$+7  kyr for  this specific  case). This  is  not very
surprising because  the dynamics becomes very  non-linear and chaotic,
the initial and numerical noise are getting amplified.

\subsection{Results for low and high thermal support}

\subsubsection{Results for a least prone to fragment case: $\alpha = 0.65$, $\beta=0.04$}

This second series of calculations is the least prone to fragmentation
because  of  its  strong  thermal  support.   We  use  a  perturbation
amplitude $A=0.5$ in  order to make fragmentation easier  if it should
occur.

\begin{figure}[h]
  \centering
  \includegraphics[width=8.5cm,height=13cm]{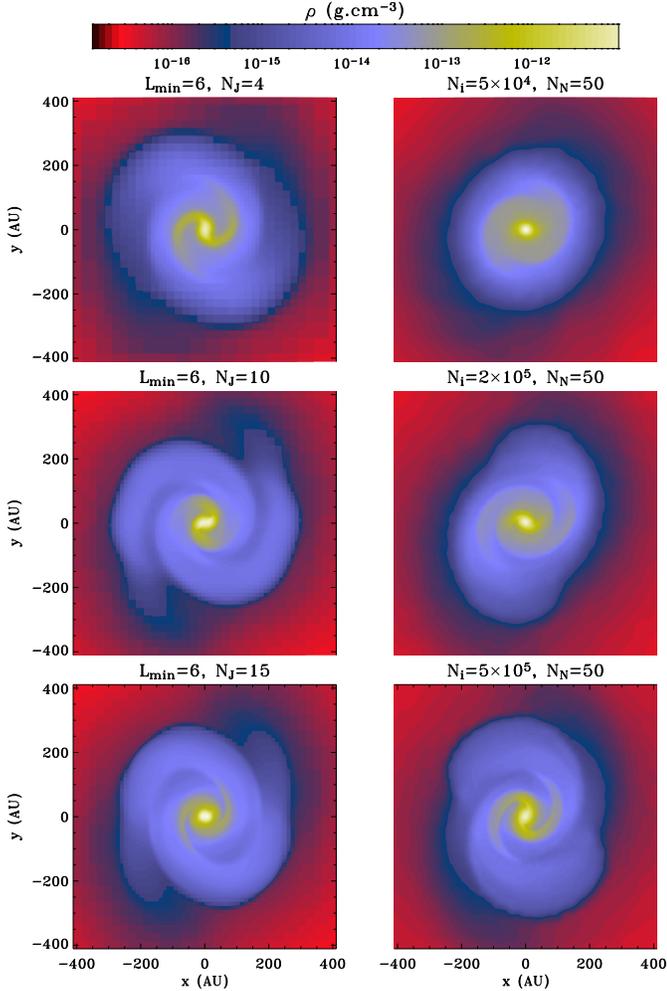}
  \caption{Density maps in the equatorial  plane at t$_0 + 10$ kyr for
$\alpha=0.65$, $\beta=0.04$  and $A=0.5$.   On the left-hand  side, we
show AMR  results with $\ell_\mathrm{min}=6$ and, from  top to bottom,
$N_\mathrm{J}=4$, 10 and 15.  The  right-hand side gives slices of SPH
calculations  with  $N_\mathrm{N}=50$ and  $N_\mathrm{p}=5\times10^4$,
$2\times10^5$ and $5\times10^5$, from top to bottom.}
\label{a065}
\end{figure}

Figure  \ref{a065} gives  density slices  on the  equatorial  plane at
t=t$_0 +  10$ kyr. On the left-hand  side, we show AMR  results for an
initial sphere described with  $\ell_{min}=6$ and, from top to bottom,
$N_\mathrm{J}=4$, 10  and 15.  The  right-hand column shows  slices of
SPH        calculations        with       $N_\mathrm{N}=50$        and
$N_\mathrm{p}=5\times10^4$, $2\times10^5$  and $5\times10^5$, from top
to bottom.   In the  previous cases, the  core has  already fragmented
into  three clumps  at this  time. In  this case,  the  cloud develops
spiral arms with no fragmentation.   The cloud fragments in some cases
after t$_0 + 50$ kyr.

AMR  and SPH  calculations converge  quickly  to a  pattern with  only
spiral  arms  and  the  formation  of a  single  central  object  when
resolution is sufficient.

\subsubsection{Early fragmentation case: $\alpha = 0.35$ , $\beta=0.04$}

This last case is the most prone to fragmentation because of its small
thermal  support  against  gravitational   energy.   In  this  set  of
calculations, the initial parameters are: $\rho_0=3.92\times 10^{-18}$
g.cm$^{-3}$,   $R_0=4.95\times   10^{16}$   cm,   $\Omega_0=3.63\times
10^{-13}$ rad.s$^{-1}$ and t$_\mathrm{ff}=1.06\times 10^{12}$ s ($\sim
33.6$  kyr).   The  initial  perturbation amplitude  is  $A=0.1$.   We
performed SPH calculations with  $N_\mathrm{p}$ ranging from $5 \times
10^4$ to $5 \times  10^5$ and $N_\mathrm{N}=50$.  The AMR calculations
were performed  with $\ell_\mathrm{min}  =6$ and 7  and $N_\mathrm{J}$
varying between  4 and  15. Although all  results are very  similar at
t$_0$, we  find some  differences at  t$_0 + 1$  kyr. For  example, it
appears quite clearly that AMR calculations with $\ell_\mathrm{min}=6$
and  $N_\mathrm{J}=4$ diverge  from the  other AMR  calculations (tiny
spiral arms).

\begin{figure}[h]
  \centering
  \includegraphics[width=8.5cm,height=13cm]{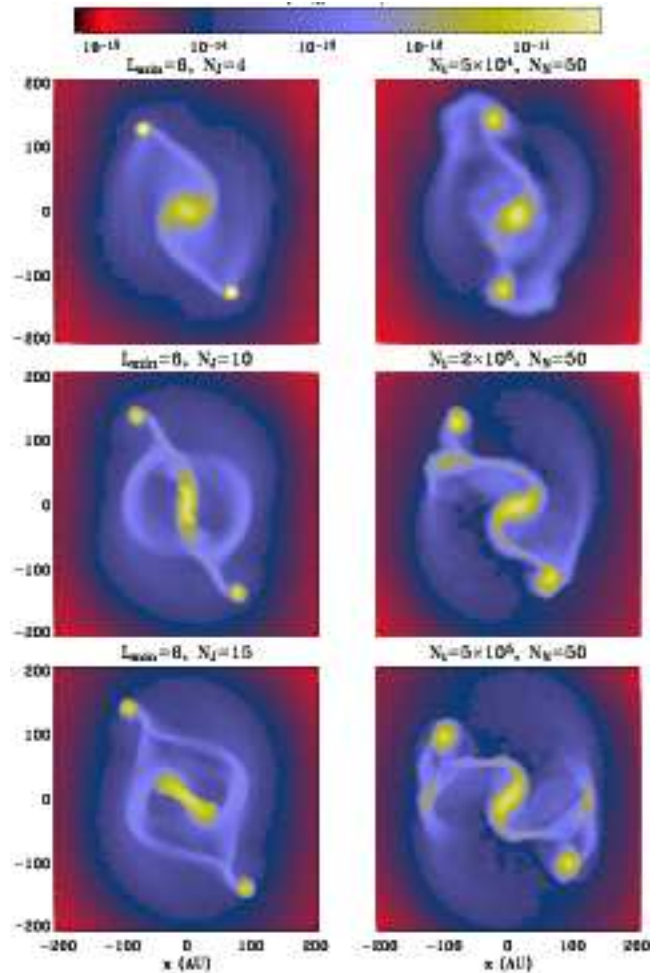}

  \caption{Density maps in the equatorial plane at t=t$_0$ + 3 kyr for
the case $\alpha=0.35$, $\beta=0.04$ and $A=0.1$. On the left side, we
give AMR  results with $\ell_\mathrm{min}=6$ and, from  top to bottom,
$N_\mathrm{J}=4$, 10 and 15. On the  right side, we show SPH maps with
calculations   with  $N_\mathrm{N}=50$  and,   from  top   to  bottom,
$N_\mathrm{p}=5\times10^4$,       $N_\mathrm{p}=2\times10^5$       and
$N_\mathrm{p}=5\times10^5$.  }
\label{a035_t4}
\end{figure}

Figure \ref{a035_t4}  shows density maps  in the equatorial  plane for
the most relevant calculations at  t=t$_0 +3$ kyr. On the left column,
we  give   the  AMR  results  with   increasing  resolution  parameter
$N_\mathrm{J}$    from    top     to    bottom    and    a    constant
$\ell_\mathrm{min}=6$. According  to our previous  results, an initial
computational domain with $\ell_\mathrm{min}=6$ is sufficient to reach
convergence  for the  AMR  calculations. The  right  column shows  SPH
calculations,  with  $N_\mathrm{p}$  ranging from  $5\times10^{4}$  to
$5\times  10^5$.  We  seem  to  reach a  convergence  between the  AMR
calculations with $\ell_\mathrm{min}=6$ and $N_\mathrm{J}>6$.  The SPH
calculations with $N_\mathrm{p}=5\times 10^4$ diverge quickly compared
to  two more  resolved  with $N_\mathrm{p}=\times  10^2$ and  $5\times
10^5$.  The most resolved AMR  and SPH runs show a convergence towards
a similar  solution.  The  patterns have the  same size  and position.
The core fragments into a central clump (of size $\sim 30$ AU) and two
identical outlying clumps (of size $\sim 10$ AU) for AMR calculations.
SPH results give a similar central object, but the outlying clumps are
larger.

However,  as shown  in Appendix  C, higher  resolution runs  show that
convergence has not  been reached.  In that case,  one needs either an
even better  resolution or,  alternatively, a more  powerful numerical
scheme.

To conclude our study on core  fragmentation, we can say that the more
non linear is  the issue, the more difficult it  is to get convergence
between SPH and  AMR simulations.  Good convergence is  found for high
enough   thermal   support.    However,   for  low   thermal   support
(i.e. $\alpha=0.35$),  convergence is more difficult  to achieve.  The
horizon of predictability in such a case is very short.

\section{Summary and Discussion}

We have investigated the effect of numerical resolution in AMR and SPH
calculations on the collapse and the fragmentation of rotating cores.
 
We show  that we  reach good convergence  between AMR and  SPH methods
provided one  uses sufficient numerical  resources.  First, we  take a
simple model to study local angular momentum conservation. The initial
study shows that  local angular momentum is better  conserved with the
AMR  approach for  equivalent computational  needs, whereas  SPH gives
better dynamical  times.  As shown in Fig.   \ref{homorot}a, a smaller
number of  particles in standard  SPH calculations leads to  bad local
angular  momentum conservation.  Numerical  torques on  the rotational
axis are  accumulated for denser  particles, whereas our  model should
remain   axisymmetric.    In   AMR   calculations,  a   poor   initial
computational domain resolution  (i.e.  $\ell_\mathrm{min} < 6$) leads
to unphysical  transfer of  gravitational energy to  rotational energy
(see Fig.   \ref{homorot}b).  A  significant loss of  angular momentum
will affect  fragmentation since  less rotational support  can balance
gravitational   collapse.   The   smallest  parameter   set   for  SPH
calculations  required to  go through  gravitational  collapse without
significant loss of angular momentum  corresponds to a number of $\sim
530$   particles   per   Jeans    mass   at   the   critical   density
$\rho_\mathrm{c}$,  i.e.  5  particles  per Jeans  {\it length}.   The
equivalent  minimum  resolution  criterion  for  AMR  calculations  is
$\ell_\mathrm{min} \ga 6$ and $N_\mathrm{J}=4$.

Then we  investigate fragmentation issues for  three different initial
condition.   For  the least  prone  to  fragment case  ($\alpha=0.65$,
$\beta=0.04$), we show  that AMR and SPH methods  give similar results
when  the respective  Jeans resolution  criteria are  fulfilled. These
results  agree  with  the  semi-analytical criteria  on  fragmentation
derived  by \cite{Tsuribe_Inutsuka_1999}  for the  isothermal collapse
($\alpha  \la 0.55  - 0.65\,  \beta$). We  study extensively  the case
$\alpha=0.5$, $\beta=0.04$.  We first reach good agreement between AMR
calculations    with   the   parameters    $\ell_\mathrm{min}=6$   and
$N_\mathrm{J}=15$  and  SPH  calculations  with  $N_\mathrm{p}=5\times
10^5$  and $N_\mathrm{N}=50$,  i.e.  $\sim  5370$ particles  per Jeans
mass at critical density $\rho_\mathrm{c}$.  These parameter sets seem
to  be  a   lower  resolution  limit  for  dense   core  collapse  and
fragmentation SPH and AMR calculations  in order to get good agreement
in both time  and space scales.  Using a lower  number of particles or
number  of points  per  Jeans  length will  lead  to inaccurate  early
fragmentation due to numerical effects.  Initially, we compare the two
converged  calculations and,  for  this specific  case,  we find  good
agreement  between the  two  methods (see  Fig \ref{comp_a050}).   The
price to pay in computer time,  however, is larger with the SPH method
for the fragmentation study, due  to not using sink particles.  In the
case  of  low  thermal  support,  the  dynamic  quickly  becomes  very
non-linear  and numerical  convergence of  the simulation  can  not be
achieved  as easily  as  for higher  thermal  support.  A  statistical
analysis over a large number of  simulations would be needed to see if
converging results  can be obtained in term  of fragment distributions
(in mass, size ...).

The   two   approaches   show   good   agreement   for   the   general
pictures. Details  are better resolved  in AMR calculations  thanks to
the refinement  method based  on the local  Jeans length,  whereas the
resolution  deteriorates with  increasing density  with  standard SPH.
Numerical  calculations  of   protostellar  collapse  should  thus  be
conducted with  great care, with  a detailed examination  of numerical
resolution. The  present work  can be used  to assess the  validity of
numerical tools to study star formation.

\begin{acknowledgements}
Calculations have been performed thanks  at the PSMN (ENS Lyon), IDRIS
and CCRT  (CEA) supercomputating facilities,  as well as on  the CEMAG
computing facility  supported by the  French ministry of  research and
education through a Chaire  d'Excellence awarded to Steven Balbus.  We
greatly  thank Anthony  Whitworth,  the referee,  for useful  comments
which have  improved the original manuscript.  We  thank Simon Goodwin
for providing the last version of the DRAGON code.
\end{acknowledgements}
\bibliographystyle{aa}

\bibliography{papier-14-01}

\begin{appendix}

\section{Complementary results on the effect of $N_\mathrm{N}$ for SPH calculations for the case: $\alpha=0.5 , \beta=0.04$ and A=0.1}\label{note_Nn}

\begin{figure}[h]
  \centering
  \includegraphics[width=8.5cm,height=18.5cm]{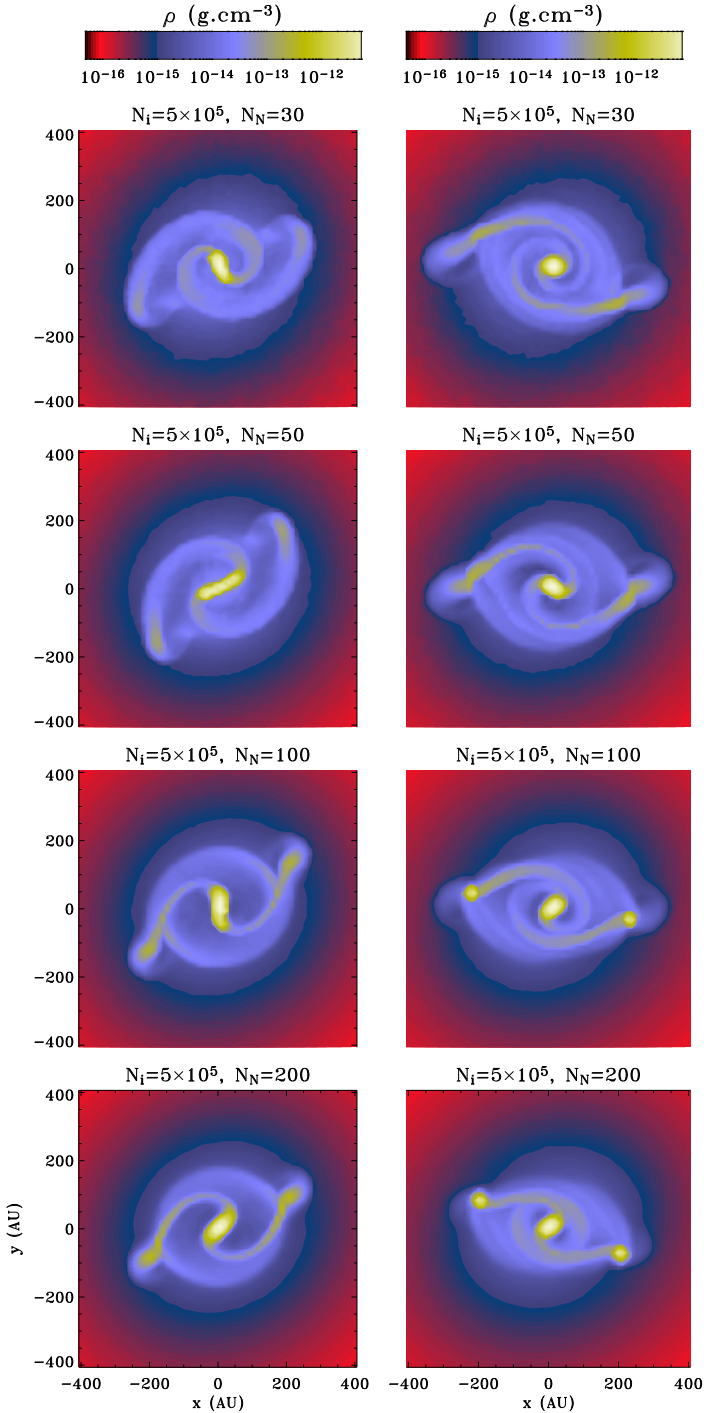}
  \caption{Density maps in the equatorial plane at two different
times from SPH calculations with $\alpha=0.50$, $\beta=0.04$ and
$N_\mathrm{p}=5\times 10^5$. The left column shows density maps for
calculations with $N_\mathrm{N}=30$, 50, 
100 and 200 from top to bottom, respectively, at t=t$_0$ + 5 kyr. The
right column represents the same calculations at t=t$_0$ + 6 kyr.}
\label{sph_nn_a050}
\end{figure}

Figure \ref{sph_nn_a050}  shows SPH  calculations run with  a constant
$N_\mathrm{p}=5\times 10^5$ and values  of $N_\mathrm{N}=$ 30, 50, 100
and 200 from top to bottom  for two timesteps (t$_0$+5 kyr on the left
column and t$_0$+6 kyr on  the right column). These simulations should
be  compared with  simulations presented  in Fig.  \ref{sph_a050}. The
first  relevant  result is  the  fact  that increasing  $N_\mathrm{N}$
speeds  up fragmentation.  Moreover, there  seems to  be  a similarity
between calculations with low $N_\mathrm{p}/ N_\mathrm{N}$ ratio, i.e.
$N_\mathrm{N}=50$,  $N_\mathrm{p}=2\times   10^5$  on  one   side  and
$N_\mathrm{N}=100$  and  $N_\mathrm{p}=5\times   10^5$  on  the  other
side. We  find the  same patterns at  different times,  postponed when
either  $N_\mathrm{p}$  increases  or  $N_\mathrm{N}$  decreases,  the
number  of  resolution elements  being  equal.   This illustrates  the
compromise between  resolution and convergence that  must be respected
in SPH calculations \citep{Lombardi_1999,rasio-1999}.

\section{Note on the artificial viscosity and numerical diffusion in SPH}\label{app_visc}
Diffusivity is a  well-known drawback of standard SPH.  This issue can
be reduced using a  constant number neighbors, $\Delta N_\mathrm{N}=0$
\citep{Attwood_2007},  and  advanced  scheme  for  viscosity  such  as
time-dependent  viscosity  \citep{Morris_Monaghan_1997}.   We  present
here SPH  calculations of the  collapse of the  uniform-density sphere
already  studied  in  \S\ref{sec_momang},  but  using  another  scheme
viscosity and/or a constant number of neighbors. This two improvements
are  quite  easy to  implement  and  do  not require  expensive  extra
computational costs.  Time-dependent  viscosity calculations have been
done with $\alpha^\star=0.1$, and an e-folding constant equal to 0.15.

\begin{figure}[htb]
 \centering
  \includegraphics[width=8cm,height=5cm]{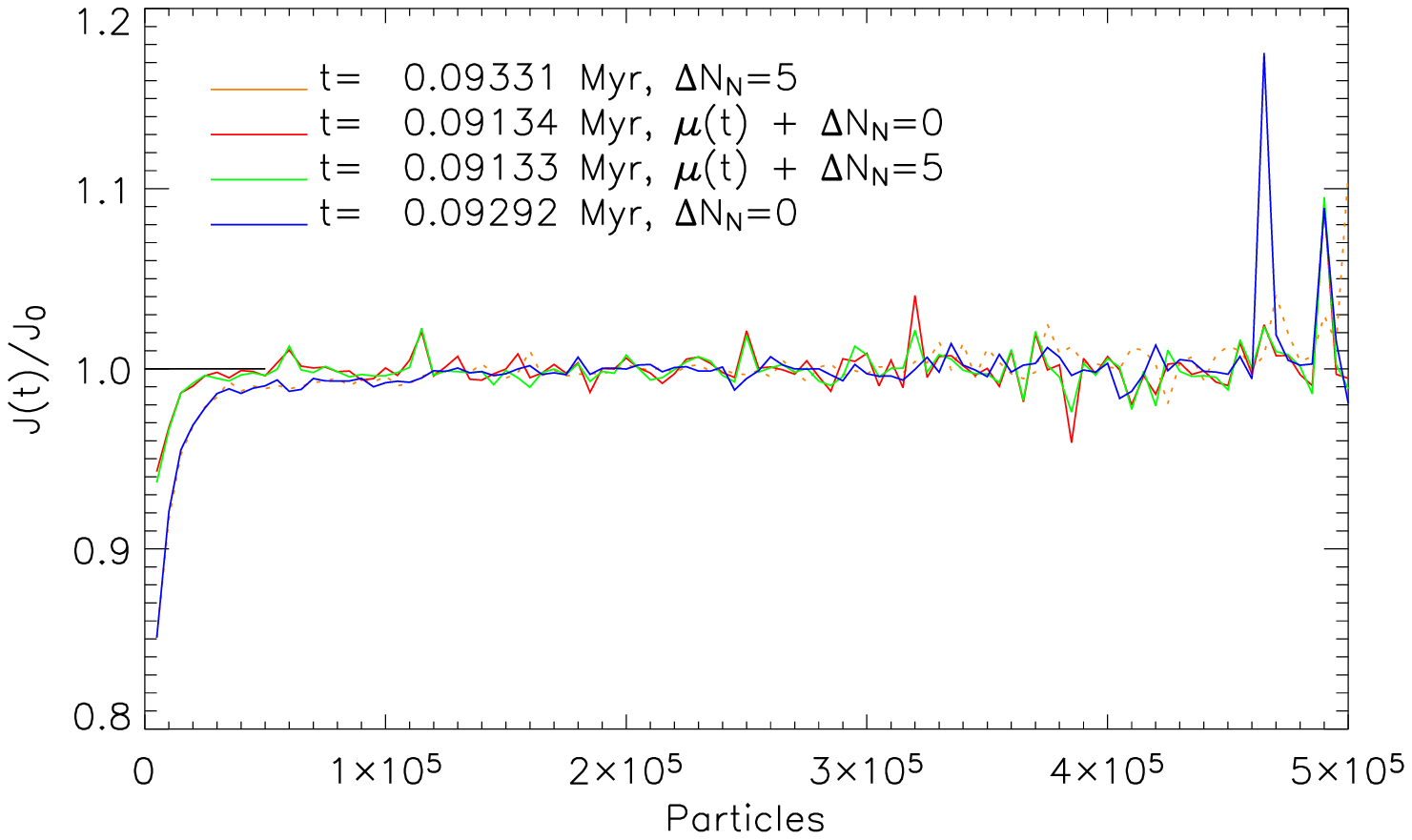}
  \includegraphics[width=8cm,height=6cm]{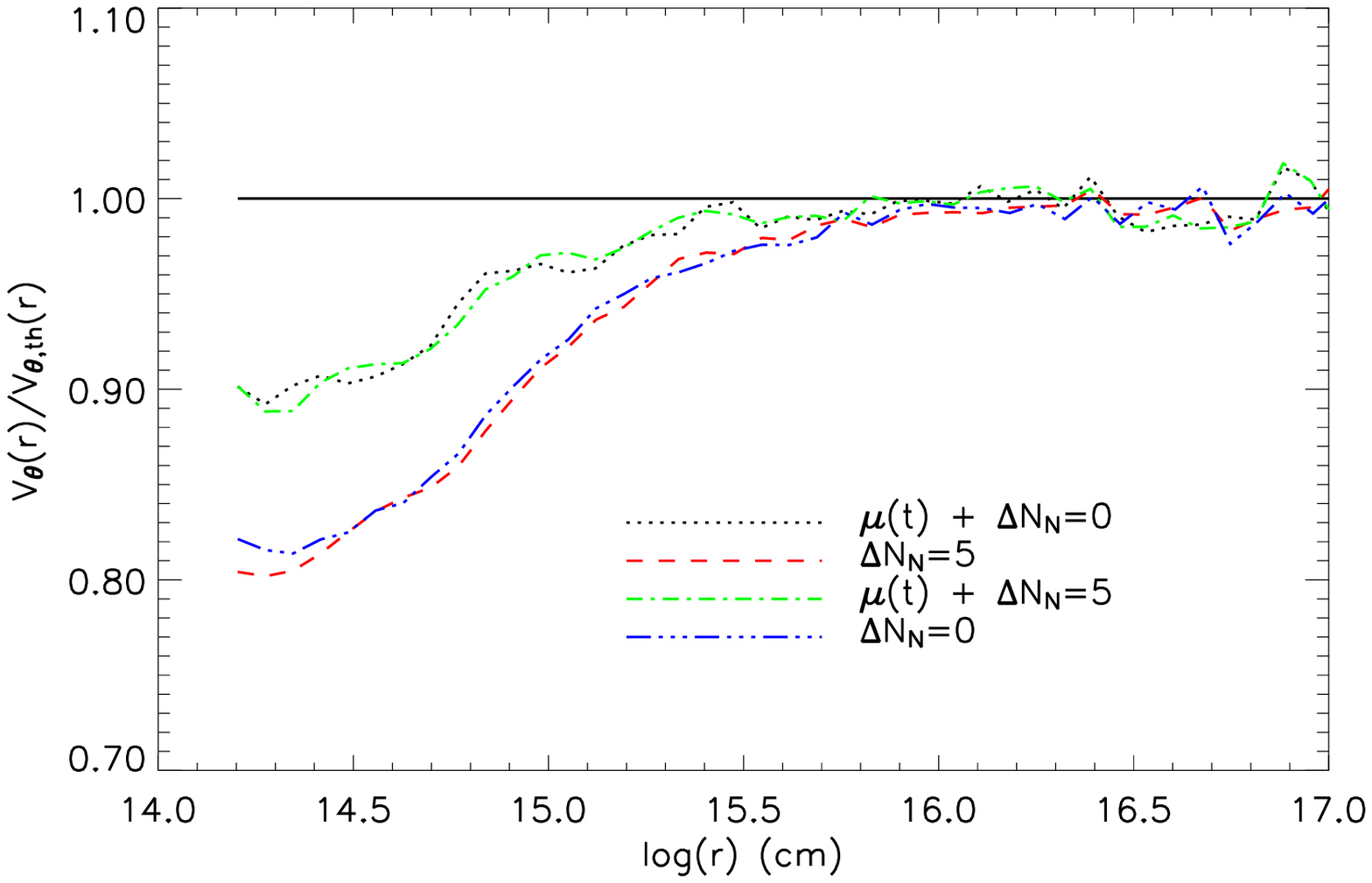}
  \caption{\it   Top  plot  \rm   (Fig  .\ref{plot_visc}a):   same  as
Fig. \ref{mom_sph} with different $\Delta N_\mathrm{N}$ and artificial
viscosity  scheme.   $\mu(t)$  indicates  the  use  of  time-dependent
viscosity instead  of the standard  artificial viscosity scheme  in the
paper.   \it   Bottom  plot  \rm  (Fig   .\ref{plot_visc}b):  same  as
Fig. \ref{mom_trans} for the above mentioned SPH calculations.}
\label{plot_visc}
\end{figure}

Figures  \ref{plot_visc}a  shows the  averaged  ratio between  angular
momentum $J(t)$ at time t$_0$  and initial angular momentum $J_0$ as a
function  of  particles  (ordered   in  decreasing  density)  for  SPH
calculations run with $N_\mathrm{p}=5\times10^5$ and $N_\mathrm{N}=50$
and the  improvement above mentioned either  turned on or  not.  It is
clear that  time-dependent viscosity better  conserve angular momentum
for the denser particles.  Hence, less angular momentum is transported
to  the  outer   part  of  the  core.   This   is  confirmed  in  Fig.
\ref{plot_visc}b where  we plot the ratio between  angular momentum at
time t$_0$  and initial angular momentum  as a function  of the radius
for  the same  calculations.  Local  angular momentum  conservation is
increased by $10\%$ in the  inner part.  However, keeping constant the
number  of   neighbors  does   not  improve  local   angular  momentum
conservation since  the system only  evolves over about  one free-fall
time  whereas  \cite{Attwood_2007}   shows  that  dissipation  becomes
significant  after a few  free-fall times.  All these  improvements of
standard SPH are as many new features that will strengthen convergence
with the AMR, particle-splitting being the most promising one.

\section{Note on the diffusion of the numerical schemes in AMR}\label{note_diff}

A key  ingredient in the  AMR method is  the numerical scheme  used to
compute  flux  at the  grid's  interfaces. In  this  paper,  we use  a
Lax-Friedrich   (hereafter  LF)  Riemann   solver  designed   for  MHD
calculations \citep{Fromang_2006}.  However, the LF scheme is known to
be a diffusive  scheme. In this appendix, we  present AMR calculations
carried out with, on one hand, the LF scheme and, on the other hand, a
Roe scheme. Roe  scheme being less diffusive than  LF, this could have
dramatic effect on the fragmentation issue.

\begin{figure}[h]
  \centering
  \includegraphics[width=8cm,height=10.67cm]{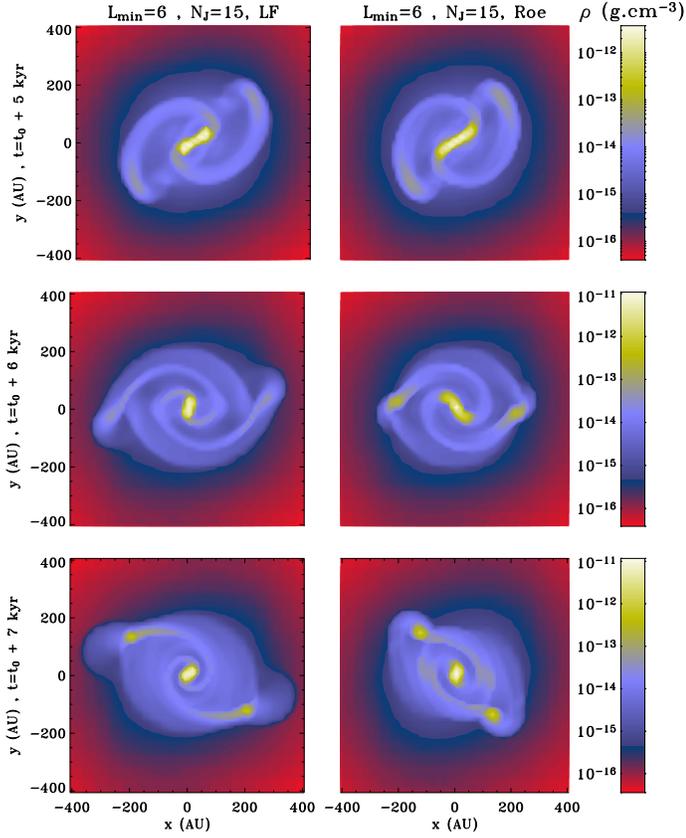}
  \caption{Density  maps   in  the  equatorial  plane   for  the  case
$\alpha=0.5$, $\beta=0.04$  at, from top  to bottom, t=t$_0$ +  5 kyr,
t=t$_0$  + 6  kyr and  t=t$_0$ +  7 kyr.  On the  left-hand  side, AMR
results     with    Lax-Friedrich    solver     and    $\ell_{min}=6$,
$N_\mathrm{J}=15$ are reported and AMR results of calculations with he
same parameters but with a Roe solver are given on the right column.
}
\label{solver_a050}
\end{figure}

\subsection{Case $\alpha=0.5$, $\beta=0.04$}

Figure \ref{solver_a050} shows density maps on the equatorial at three
different  timesteps  for  two  AMR  calculations run  with  the  same
numerical parameters,  i.e. $\ell_{min}=6$ and  $N_\mathrm{J}=15$, but
with a  different solver, i.e. the LF  one on the left  column and the
Roe  one  on  the  right  column.   Results  are  quite  similar,  AMR
calculations are in good agreement for this critical case with the two
solvers.   Since  less  angular  momentum  has been  locally  lost  or
transported with the Roe scheme, the core is smaller and the fragments
are closer  to the  central object.  This  brings support to  the fact
that we  find good  convergence between AMR  and SPH  calculations for
this case.

\subsection{Case $\alpha=0.35$, $\beta=0.04$}

\begin{figure}[h]
  \centering
  \includegraphics[width=8cm,height=13.33cm]{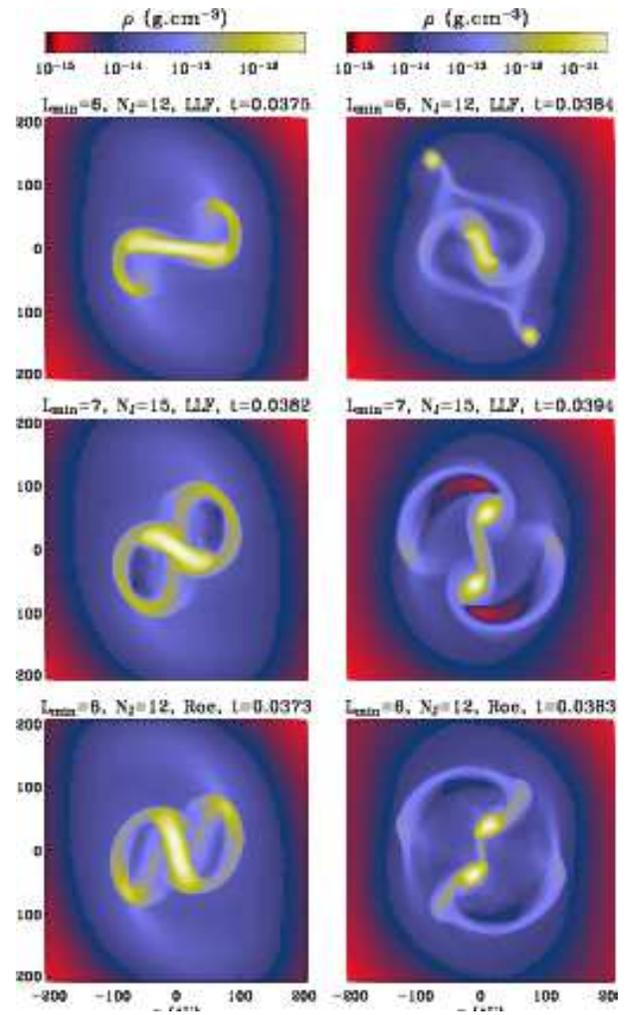}
  \caption{Density  maps   in  the  equatorial  plane   for  the  case
$\alpha=0.35$, $\beta=0.04$ at  t=t$_0$ + 2 kyr on  the left-hand side
and t=t$_0$ + 3 kyr on the right-hand side. For the two upper rows, we
plot  results   for  AMR  calculations   with  ($\ell_\mathrm{min}=6$,
$N_\mathrm{J}=12$), ($\ell_\mathrm{min}=7$, $N_\mathrm{J}=15$) and our
usual  Lax-Friedrich scheme.   The bottom  row gives  the  results for
calculations conducted with ($\ell_\mathrm{min}=6$, $N_\mathrm{J}=12$)
too, but with a Roe solver.  Times are given in Myr.}
\label{solver}
\end{figure}

 Figure  \ref{solver} shows density  maps on  the equatorial  plane at
t$_0$ + 2  kyr (right column) and t$_0$ + 3  kyr for three simulations
of the  case $\alpha =  0.35$, $\beta=0.04$ with  numerical parameters
$\ell_\mathrm{min}=6$ and  $N_\mathrm{J}=12$ (top and  bottom maps, LF
and  Roe  schemes)  and  $\ell_\mathrm{min}=7$  and  $N_\mathrm{J}=15$
(middle row, LF  scheme).  Let us remind that  in Fig.  \ref{a035_t4},
the  case  $\ell_\mathrm{min}=6$  and  $N_\mathrm{J}=15$ with  the  LF
solver  has  been  displayed.   The  two calculations  with  the  same
numerical parameters  differ, according to the  numerical scheme used.
The fragmentation  process changes:  one gets a  configuration central
object +  two satellites with  the LF scheme  whereas we get  a binary
system resulting from the fragmentation of the central object with the
Roe  scheme.  If  we  improve  the initial  sphere  resolution  in  LF
calculations  (i.e.    $\ell_\mathrm{min}=7$,  $N_\mathrm{J}=15$),  we
converge to the results obtained  with the Roe scheme, i.e.  a central
binary  system,  with a  value  $\ell_\mathrm{min}<7$.   We know  that
angular  momentum is well  conserved using  the Roe  scheme or  the LF
scheme with $\ell_\mathrm{min}=7$, so  it seems that calculations lead
to  a different  core  fragmentation because  of  their less  accurate
angular momentum conservation.  Since  we use a small thermal support,
it  is  easy  to  reach  another  fragmentation  configuration,  these
processes being highly non-linear.

\section{Note on SPH sink particles}\label{note_sink}

The  introduction of  sink particles  is a  widely used  way to  get a
compromise  between good  resolution  and acceptable  timestep in  SPH
methods.  Creating  a sink  particle  enables  to  loosen the  Courant
condition on the particle timesteps.

The  density level $\rho_\mathrm{sink}$  at which  a sink  particle is
created has to be chosen  with care. In the previous SPH calculations,
no sink  particles were  used. Let us  focus on the  highly non-linear
fragmentation case, $\alpha=0.35$, $\beta=0.04$ and $A=0.1$ to present
SPH calculations carried out with various sink densities.

\begin{figure}[h]
  \centering
  \includegraphics[width=8cm,height=13.33cm]{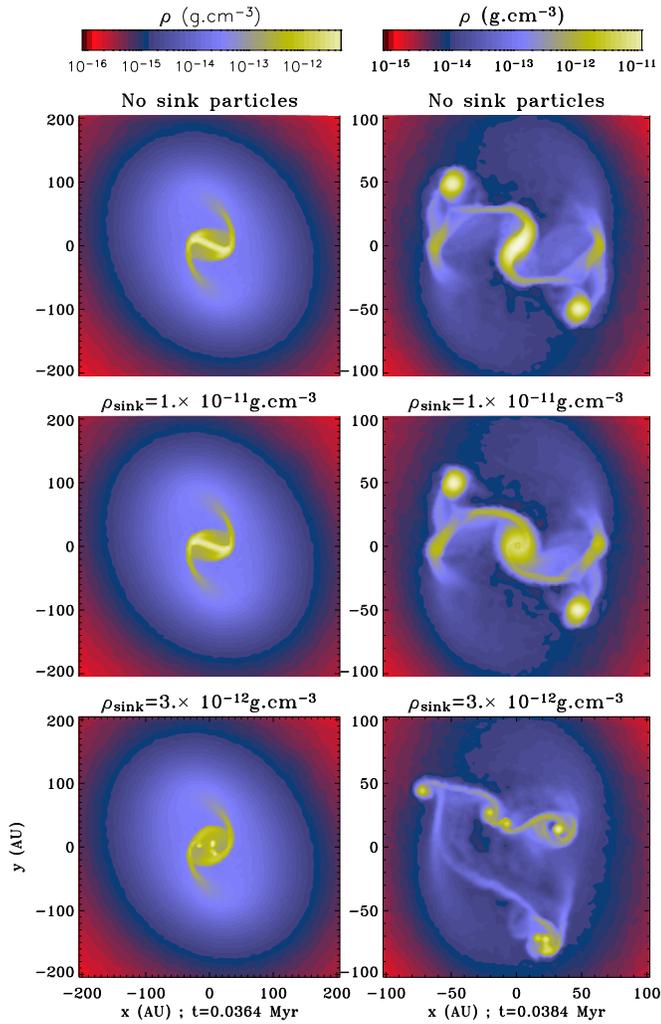}
  \caption{Density  maps   in  the  equatorial  plane   for  the  case
$\alpha=0.35$, $\beta=0.04$ at  t=t$_0$ + 1 kyr on  the left-hand side
and t=t$_0$ + 3 kyr on the right-hand side. 9 sink particles have been
created  on the  bottom figures,  affecting the  whole dynamic  of the
dense core.}
\label{sink}
\end{figure}

Figure \ref{sink} shows calculations  carried out with three different
densities  for the creation  of sink  particles, $\rho_\mathrm{sink}$,
namely $1\times 10^{-10}$ g.cm$^{-3}$ (resulting in no sink creation),
$1\times  10^{-11}$ g.cm$^{-3}$  and $3\times  10^{-12}$ g.cm$^{-12}$,
from top to bottom, and the same number of SPH particles and neighbors,
$N_\mathrm{p}=5\times 10^5$ and  $N_\mathrm{N}=50$. The left-hand side
reports results at  t$_0$+ 1 kyr and the  right-hand column results at
t$_0$  + 3  kyr.  The CPU  time is  about  28\% smaller  for the  case
$\rho_\mathrm{sink}=3\times10^{-12}$  g.cm$^{-3}$  than  for the  case
$\rho_\mathrm{sink}=1\times10^{-10}$  g.cm$^{-3}$, but the  dense core
resulting  at  t$_0  + 3$  kyr  is  really  different. The  upper  row
corresponds  to  calculations  without  creation  of  sink  particles,
whereas     one    sink    particle     has    been     created   with
$\rho_\mathrm{sink}=1\times10^{-11}$    g.cm$^{-3}$    and    9   with
$\rho_\mathrm{sink}=3\times10^{-12}$  g.cm$^{-3}$. It  is easy  to see
that, even if only one  sink particle is created, the complete dynamic
is affected particularly in the central region.

In conclusion, it is clear  that sink particles should be handled with
great  care.   A  fair   comparison  should  also  compare  these  SPH
calculations with AMR ones  including sink particles. Such studies are
in progress.

\end{appendix}

\end{document}